\documentclass[usenatbib]{mnras}
\usepackage{newtxtext,newtxmath}

\usepackage{natbib}
\usepackage{longtable}
\usepackage{placeins}
\usepackage{graphicx}	
\title[Search for the optical counterpart of GW170814 ]{Search for the optical counterpart of the GW170814 gravitational wave event with the VLT Survey Telescope}
\author[A. Grado ]
{A. Grado$^{1,2}$\thanks{\vspace {-1.5cm} E-mail: aniello.grado@inaf.it},
E. Cappellaro$^3$,
S. Covino$^4$,
F. Getman$^1$,
G. Greco$^5$,
L. Limatola$^1$,
\newauthor
S. Yang$^3$,
L. Amati$^6$,
S. Benetti$^3$,
M. Branchesi$^{7,8}$,
E. Brocato$^{9,12}$,
M. Botticella$^1$, 
\newauthor
S. Campana$^4$,
M. Cantiello$^{9}$,
M. Dadina$^6$,
F. D'Ammando$^{10}$,
G. De Cesare$^6$,
\newauthor
V. D'Elia$^{12,11}$,
M. Della Valle$^1$,
E. Iodice$^1$,
G. Longo$^{13}$,
M. Mapelli$^{14}$,
N. Masetti$^{6,15}$,
\newauthor
L. Nicastro$^6$,
E. Palazzi$^6$,
A. Possenti$^{16}$,
M. Radovich$^3$,
A. Rossi$^{6,11}$,
R. Salvaterra$^{17}$,
\newauthor
L. Stella$^{12}$,
G. Stratta$^{6,18}$,
V. Testa$^{11}$,
L. Tomasella$^3$,
\vspace{0.3cm}
\\
$^{1}$ INAF, Osservatorio Astronomico di Capodimonte, Salita Moiariello 16, I-80131, Napoli, Italy\\
$^{2}$ INFN, - Sezione di Napoli - Complesso Universitario di M. S. Angelo, Ed. 6- Via Cintia, I-80126 Napoli, Italy\\
$^{3}$ INAF, Osservatorio Astronomico di Padova, Vicolo dell'Osservatorio 5, I-35122 Padova, Italy\\
$^{4}$ INAF, Osservatorio Astronomico di Brera, Via E. Bianchi 46, I-23807 Merate (LC), Italy\\
$^{5}$ Universit\`a degli Studi di Urbino `Carlo Bo', Dipartimento di Scienze Pure e Applicate, P.za Repubblica 13, I-61029, Urbino, Italy\\
$^{6}$ INAF - Osservatorio di Astrofisica e Scienza dello Spazio, Via Piero Gobetti 93/3, I-40129 Bologna, Italy\\
$^{7}$ Gran Sasso Science Institute, Viale F. Crispi 7, I-67100 L'Aquila, Italy\\
$^{8}$ INFN, Laboratori Nazionali del Gran Sasso, I-67100 Assergi, Italy\\
$^{9}$ INAF, Osservatorio Astronomico d'Abruzzo, Via Mentore Maggini
Teramo, TE, I-64100, Italy\\
$^{10}$ INAF, Istituto di Radioastronomia, Via Piero Gobetti 101, I-40129 
Bologna, Italy\\
$^{11}$ ASI Space Science Data Centre, Via del Politecnico snc, I-00133, Roma, Italy\\
$^{12}$ INAF, Osservatorio Astronomico di Roma, Via Frascati, 33, I-00040 Monte Porzio Catone, Roma, Italy\\
$^{13}$ Universit\`a degli Studi di Napoli Federico II, Complesso Universitario di Monte Sant'Angelo, Via Cinthia, 21 Edificio 6, I-80126, Napoli, Italy\\
$^{14}$ Physics and Astronomy Department "G. Galilei"
University of Padova, Vicolo dell'Osservatorio 3
I-35122 Padova, Italy\\
$^{15}$ Departamento de Ciencias F\'isicas, Universidad Andr\'es Bello, Fernandez Concha 700, Las Condes, Santiago, Chile\\
$^{16}$ INAF, Osservatorio Astronomico di Cagliari, Via della Scienza 5, I-09047 Selargius (CA), Italy\\
$^{17}$ INAF/IASF-MI, via Bassini 15, I-20133 Milano, Italy\\
$^{18}$ INFN-Firenze, via Sansone 1, I-50019, Firenze, Italy\\
}
\date{Accepted ... Received ...; in original form ...}
\pubyear{2019}
\begin{document}
\maketitle
\begin{abstract}
We report on the search for the optical counterpart of  the gravitational event GW170814, which was carried out with the VLT Survey Telescope (VST) by the GRAvitational Wave Inaf TeAm (GRAWITA). Observations started 17.5 hours after the LIGO/Virgo alert and we covered an area of  99 deg$^2$ that encloses  $\sim 77\%$ and  $\sim 59\%$ of the initial and  refined localization probability  regions, respectively. A total of six epochs were secured over nearly two months. The survey reached an  average limiting magnitude of 22 AB mag in the $r-$band.  After assuming the model described in \cite{2019ApJ...875...49P}, that 
derives as possible optical counterpart of a BBH event a transient source declining in about  one day, we have computed a survey efficiency of about $5\%$. This paper describes the VST observational strategy and the results obtained by our analysis pipelines developed to search for optical transients in multi-epoch images. We report the catalogue  of the candidates with possible  identifications based on light-curve fitting. We have identified two dozens of SNe, nine AGNs, one QSO. Nineteen transients characterized by a single detection were not classified. We have restricted our analysis only to the candidates that fall into the refined localization map. None out of 39 left candidates could be positively associated with GW170814. This result implies that the possible emission of optical radiation from a BBH merger had to be fainter than r $\sim$ 22 ($L_{optical}$ $\sim$ $1.4 \times 10^{42}$ erg/s) on a time interval ranging from a few hours up to two months after the GW event.
\end{abstract}

\begin{keywords}
gravitational wave: general --- gravitational wave: individual (GW170814); wide field surveys; VST
\end{keywords}

\section{Introduction}
August 14, 2017 is a milestone in the gravitational waves (GW) Astronomy. The Virgo interferometer \citep{Acerneseetal2015}  detected its first signal and the triangulation with the LIGO interferometers \citep{Aasietal2015}  led to a strong improvement of the sky localization of the event GW170814, shrinking the area  of the  original error box obtained by using only the LIGO detectors, from 1160 deg$^2$ to only 60 deg$^2$ after using all three detectors \citep{Abbottetal2017a}\footnote{The on-line False Alarm Rate (FAR) value is given in \cite{2019arXiv190103310T} and the final estimates of the parameters of the two black hole are given in \cite{2018arXiv181112907T}.}. This was a fundamental step forward in the newly born era of multi-messenger astrophysics which was fully exploited three days later with the detection of the binary neutron stars merger GW\,170817 event and its electromagnetic counterpart \citep{Abbottetal2017b,2017Natur.551...67P}. 

The GW170814 event was detected at 10:30:43 UTC; it is attributed to the merger of two black holes (BH) with a false-alarm rate of $1/27000$ years. The estimated black hole masses are $30.5^{+5.7}_{-3.0} M_{\odot}$ and $25.3^{+2.8}_{-4.2} M_{\odot}$, while the inferred luminosity distance is $540^{+130}_{-210}$ Mpc\footnote{We assume  a flat cosmology with Hubble parameter
$H_0$ = 67.9 km s$^{-1}$ Mpc$^{-1}$ and matter density parameter $\Omega_m$ = 0.3065 \citep{2016A&A...594A..13P}} corresponding to a redshift  $z = 0.11^{+0.03}_{-0.04}$ \citep{Abbottetal2017a}. 
In the  O1 and O2 LIGO and Virgo observing run a total of 10 BBHs have been observed  \citep{2018arXiv181112940T}, allowing us to put constraints on a certain number of parameters associated to the population of BBHs.  Among them it has been possible to infer a BBH merger rate density of $\sim$ 60 Gpc$^{-3}$yr$^{-1}$. Of high valuable scientific interest is their mass distribution, it comes out that there is a significant reduction in the merger rate for BBHs having primary masses larger that $\sim$ 45 M$_{\odot}$, while the limited sensitivity does not allow to place reliable constraints on the minimum mass of black holes. The limited size sample and the local nature of the events so far observed have prevented to infer the BBH redshift distribution. 
A binary BH merging is not expected to produce an electromagnetic (EM) counterpart. However, there are alternative models \citep{deMink&King2017, 2016ApJ...821L..18P, 2018MNRAS.477.4228P, 2016ApJ...827L..31Z, 2016ApJ...822L...9M, 2016PTEP.2016e1E01Y, 2017MNRAS.464..946S} which predict possible EM radiation in the optical/near infrared spectral region. 
 If a fraction of the progenitors mass lost during the prior evolution of the binary stellar system,  is retained as circumbinary disk, an infrared emission to medium-energy X-rays  might be observable within hours after the GW event \citep{deMink&King2017}.
In particular \cite{2019ApJ...875...49P} find
that after one day from the GW event an optical signal in the R band could be detected at a magnitude of $\sim$22.5 for a source at a distance of GW150914.

The detection of an EM counterpart of a GW emitted by a stellar BBH  would reduce GW parameters degeneracy (see \cite{2017ApJ...834..154P}), it would allow the measurement of the source redshift \citep{1986Natur.323..310S}, and provide hints on the formation channel of the BBH binary by exploring the environment in which the merger occurred. Indeed dynamical formation \citep{2016PhRvD..93h4029R} is likely to take place in dense star clusters,  while the isolated binary evolution is expected in typical field galaxies \citep{2007PhR...442...75K}. 
The results of the search of optical counterpart to GW170814 event was recently published \citep{2019ApJ...873L..24D}. By using the Dark Energy Camera, 225 deg$^2$ were imaged corresponding to 90 per cent of the LALinference probable sky area. The observations have been carried in the i  band  reaching a depth of  $\sim$ 23 mag. With a selection criteria optimized to search for  fast declining (few days) transients two candidates were found. However, as stated by the authors, these candidates are most likely not associated with GW170814.
This work is carried out within the GRAWITA collaboration \citep{Brocatoetal2018} and is organized as follows: section 2 describes the VST observational strategy, section 3 the processing of data and search for counterpart. In section 4 is described the survey detection efficiency calculation. Finally, we draw our conclusions in section 5.

\section{VST observational strategy}
We carried out the search for the optical counterpart of the GW170814 event using the ESO VLT Survey Telescope \citep[VST,][]{Capacciolietal2003} located at Cerro Paranal in Chile. VST is a 2.6m telescope specifically designed  for wide-field imaging. It is equipped with OmegaCAM, a CCD mosaic camera made of 32 ($2048\times 4096$) pixel devices for a total $16$k $\times 16$k pixels, providing a field of view of $\sim$ 1  deg$^2$ \citep{Kuijken2011}. The standard filter set includes {\em ugriz} Sloan filters \citep{Doietal2010}.

We started our observations 17.5 hours \citep{Grecoetal2017} after the LIGO/Virgo alert  triggered by four independent low-latency  pipelines at 2017-08-14 10:30:43 UTC  \citep{GCN21474}.
The sky area tiling was obtained using a dedicated tool named \emph{GWsky}\footnote{https://github.com/ggreco77/GWsky}.
\emph{GWsky} is a {\it python}\footnote{http://www.python.org} tool developed to generate accurate sequences of pointings (tiles) covering the sky localization of a GW. It is equipped with a Graphical User Interface (GUI) optimized for fast and interactive telescope-pointing operations and supplies information and descriptive statistics on telescope visibility, GW localization probability, availability of reference images and coverage of catalogued galaxies for each FoV footprint. GWsky supports two observational strategies according to the preliminary information reported in the LIGO/Virgo GCN notices/circulars. They are  $(1)$ the tiling and  $(2)$ targeted searches. The tiling (or coverage) strategy mainly provides a list of pointings that $(i)$ optimizes the telescope's visibility during the pre-assigned epochs, $(ii)$ maximizes the contained localization probability and $(iii)$ avoids areas with too bright objects and/or severe crowding.
The second strategy generates a list of pointings covering galaxies from the Glade catalogue \citep{Dalyaetal2018} ranked  from the highest to the lowest luminosity taking into account the initial airmass.

The posterior luminosity distance and standard deviation of  GW170814\footnote{The GW trigger was recorded with LIGO-Virgo ID G298048, by which it is referred throughout the GCN Circulars.} were automatically taken from the FITS header of the initial BAYESTAR sky map \citep{2016PhRvD..93b4013S} ({\tt{DISTMEAN}} and {\tt{DISTSTD}}) for rapidly setting an observational strategy. 
Given the nature of the event and its posterior mean distance, the number of galaxies contained in the sky error volume was found to be too high for a targeted search so that the {\emph{coverage}} strategy was chosen. 
Starting from the maximum probability pixel of the BAYESTAR sky map, (RA[deg] = 41.06842, DEC[deg] = -45.48795) a sequence of 9 consecutive tiles (each tile is made of $3\times3$ pointings \footnote{The choice of 3$\times$3 $deg^2$ tiles is due to technical reasons, it optimizes the telescope overheads.}) were defined limiting the airmass to 2 and maximizing the enclosed probability in each tile. The survey was performed in the {\it r}-band covering an area of $81\,\mbox{deg}^2$, enclosing the initial sky localization probability of $\sim$ 76 per cent.
The choice of the r band is a trade off between telescope and camera efficiency and expected spectral distribution of the electromagnetic signal.

\begin{figure*}
\includegraphics[angle=0,width=0.70\textwidth]{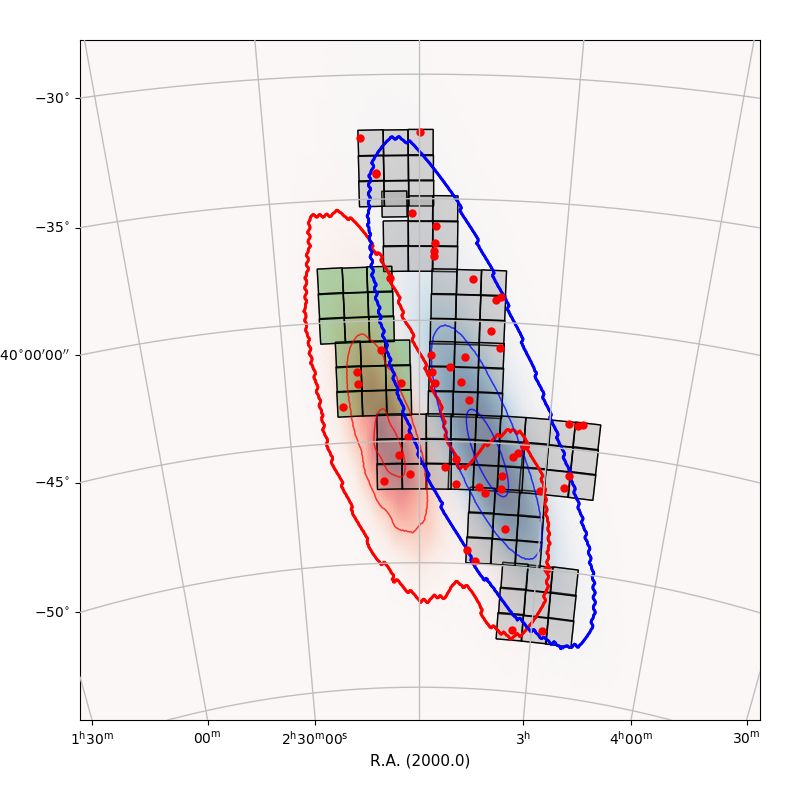}
\caption{Footprints of the VST $r-$band observations over the contours of both the initially distributed BAYESTAR  and refined LALinference sky localization maps of GW 170814. Each small square represents a VST observation. The blue and red external contours are the initial localization and refined map at 90 per cent confidence level, respectively. The inner contour lines represent the enclosed probabilities of 50 per cent  and 10 per cent confidence level, respectively. The tiles in light green are observed since the third epoch.}
\label{fig:tiles}
\end{figure*}

On 2017 August 16,  an updated localization from LIGO and Virgo data around the time of the compact binary coalescence was provided \citep{GCN21493}. Source parameters estimation were performed using the LALInference algorithm \citep{2015PhRvD..91d2003V}  and a new sky map was available for retrieval from the GraceDB\footnote{https://gracedb.ligo.org/} event page (hereafter, preliminary LALInference). 
The maximum probability pixel of the preliminary LALInference is localized at the coordinates RA[deg] = 46.53409, DEC[deg] = -44.59799. The 90 per cent credible regions of the initial (BAYESTAR) and his updated localization have significant overlap, but the 50 per cent credible region is shifted east by its entire width.
To compensate such shift,  2 additional tiles were added (shown in light green in figure \ref{fig:tiles}), for a total observed area of 99 deg$^2$, starting from the third epoch enclosing a total probability of $\sim$ 50 per cent of the Preliminary LALInference. On 2017 September 27, a refined sky localization was issued taking into account calibration and waveform modelling uncertainties \citep{GCN21934}.

The sky coordinates of the VST pointings and the enclosed probability for each tile are reported in Table~\ref{tab:prob_pointings}. The  enclosed probability is calculated with respect to the BAYESTAR  \citep{GCN21474}, LALInference preliminary \citep{GCN21493} and the final LALInference  (hereafter, refined LALInference) skymaps. The refined LALInference was recently published in the Gravitational-Wave Transient Catalogue - GWTC-1  \citep{2018arXiv181112907T}. 
The initial BAYESTAR sky localization and the refined LALInference sky localization, as well as the VST tiles, are shown in Fig. \ref{fig:tiles}.
We covered a total area of  99 deg$^2$,  taking into account uncovered gaps among ccds, haloes and spikes due to bright stars,  we encloses $\sim$ 77 per cent and  $\sim$ 59 per cent of the initial and  refined localization probability region, respectively.

The VST observations are organized in  observing blocks (OB)  consisting of a mosaic of 9  pointings -- $1 \times 1\, \mbox{deg}^2$ --  hence covering a total area of $3 \times 3\, \mbox{deg}^2$.   Each pointing is made of two exposures of 40 s each dithered by $0.7 - 1.4\,\mbox{arcmin}$. The dithering allows us to fill the gaps in the OmegaCAM CCD mosaic. 
The VST observations of the patrolled area were repeated six times distributed over $\sim$ two months in order to have enough data points to sample the light curve of the optical transients.  A 50\% point sources  completeness  at r $\simeq$  22.0 AB mag  was reached  in  most  epochs. 

\begin{figure*}
\centering
\includegraphics[angle=0,width=\textwidth]{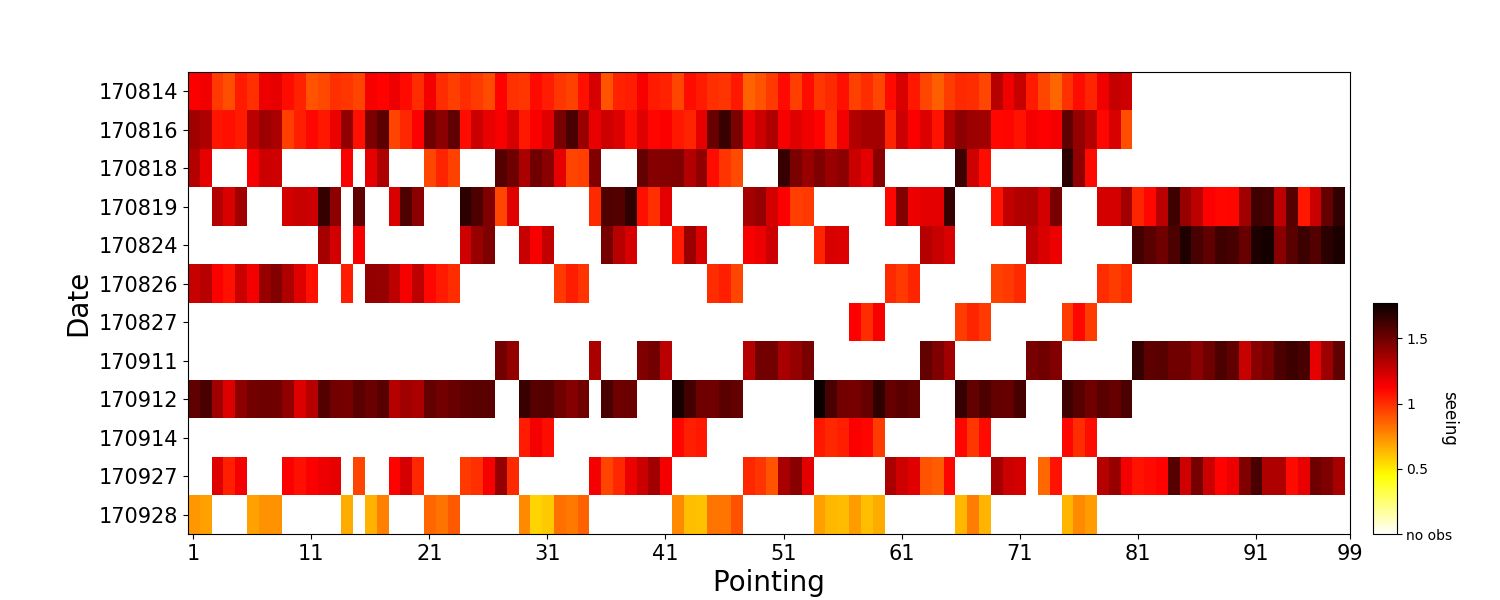}
\caption{Overview of the survey monitoring progress. On ordinates are indicate months and days. On abscissa the VST pointings of 1 $\times$ 1 deg$^2$.  The color indicates the seeing achieved for each specific pointing. Please refer to Table \ref{tab:data} to see the way nights are grouped for each epoch.}
\label{fig:monitoring}
\end{figure*}

A summary of the VST observations performed for the GW170814 event is reported in Table~\ref{tab:data}.  In some epoch it was not possible to cover the whole area in one single night. In Tab.~\ref{tab:data} we also report the average seeing (FWHM), the minimum and maximum values of the airmass, the magnitude corresponding to 50\% completeness and  the minimum and maximum value for the different pointings \footnote{The 50\% completeness is estimated using {\it VST-Lim}, a tool of the {\tt VST-Tube} pipeline \citep{Gradoetal2012}, that adds to the image under analysis simulated point sources \citep[generated using Skymaker,][]{Bertin2009} and counts for the number of recovered sources as function of the magnitude.}.
An overview of the temporal distribution of the observations along with the observing conditions is shown in Fig.\,\ref{fig:monitoring}. 

\begin{table*}
\centering
\caption{Summary of the VST tiled observations. For each pointing the contained probability is measured from the initial to the final GW170814 sky localization. The last two tiles $T_{10}$ and $T_{11}$ were observed starting from the third epoch.}
\label{tab:prob_pointings}
\begin{tabular}{lccccccc}
\\
\hline\hline
 Tile & RA & Dec &  BAYESTAR & Preliminary LALInf. &  Refined LALInf. & \\
         &  ICRSd   &  ICRSd &  Contained Probability (\%) & Contained Probability (\%) &  Contained Probability (\%) &  \\
\hline

 T$_{1}$       &02:44:16.421 &-45:29:16.62  &22.0  &2.3  &1.9  \\ 
 T$_{2}$       &02:27:09.286 &-45:29:16.62  &3.8  &0.6  &0.7    \\
 T$_{3}$       &03:01:23.556 &-45:29:16.62  &2.6  &15.0  &22.0    \\ 
 T$_{4}$       &02:49:41.880 &-42:29:16.62  &17.0  &1.5  &0.5    \\ 
 T$_{5}$       &02:49:41.880 &-39:29:16.62  &7.1  &0.4  &0.08   \\ 
 T$_{6}$       &02:38:50.962 &-48:29:16.62  &14.0  &3.1  &4.1   \\ 
 T$_{7}$       &02:59:53.815 &-36:29:16.62  &3.7  &0.5  &0.1   \\ 
 T$_{8}$       &03:04:41.582 &-33:48:16.60  &1.5  &0.4  &0.07   \\ 
 T$_{9}$       &02:29:17.376 &-51:29:16.62  &3.8  &1.7  &1.8   \\ 
 T$_{10}$      &03:10:23.030 &-42:25:52.18  &0.7  &14.0  &20.0   \\ 
 T$_{11}$      &03:13:39.854 &-39:22:22.33  &0.6  &10.0 &7.8   \\  
\hline
\end{tabular}
\end{table*}

\begin{table*}
\centering
\caption{ Summary of the VST observations performed for the GW170814 event. One epoch is completed when the whole sky map is covered. For observational constraints not necessarily one epoch is completed in a single night as reported in column 1. The second column reports the night that includes all the observations carried out between 12:00 UT of the indicated night and finished at 12:00 UT of the day after. The third column reports the covered area in deg$^2$.  Then, it is shown the average FWHM, the minimum and maximum values of the airmass, the average 50 per cent completeness limit and its min and max values.}
\label{tab:data}
\begin{tabular}{lccccccc}
\\
\hline\hline
Epoch & Date & Area &  FWHM & AIRMASS & compl. avg & compl.\\
     &   (UT)   &    deg$^2$ & arcsec & min-max & 50\% & min-max \\
\hline
1       & 2017-08-14 & 81 & 1.04 & 1.05-1.98 & 22.0 & 21.8-22.3\\ 
2       & 2017-08-16 & 81 & 1.23 & 1.05-1.98 & 22.0 & 21.4-22.4 \\
3       & 2017-08-18 & 45 & 1.29 & 1.15-1.98 & 21.8 & 21.2-22.3 \\ 
3       & 2017-08-19 & 45 & 1.31 & 1.05-1.79 & 22.0 & 21.6-22.4 \\ 
4       & 2017-08-24 & 38 & 1.39 & 1.08-1.79 & 21.9 & 21.6-22.2\\ 
4       & 2017-08-26 & 31 & 1.12 & 1.05-1.98 & 22.2 & 22.0-22.4\\ 
4       & 2017-08-27 & 9  & 1.03 & 1.39-1.51 & 22.4 & 22.5-22.5\\ 
5       & 2017-09-11 & 36 & 1.46 & 1.08-1.25 & 21.5 & 21.2-21.9\\ 
5       & 2017-09-12 & 43 & 1.52 & 1.05-1.98 & 21.7 & 21.5-22.1\\ 
5       & 2017-09-14 & 18 & 1.07 & 1.22-1.51 & 22.2 & 22.1-22.4\\ 
6       & 2017-09-27 & 63 & 1.19 & 1.05-1.79 & 22.0 & 21.5-22.4\\ 
6       & 2017-09-28 & 36 & 0.72 & 1.22-1.98 & 22.0 & 21.6-22.3\\ 
\hline
\end{tabular}
\end{table*}

\section{Data Processing}
\label{dataproc}

\subsection{Pre-reduction}

Immediately after acquisition, the images were mirrored to the ESO Garching data archive, and then  transferred by an automatic procedure from the ESO Headquarters to the VST Data Center in Naples (Italy). The first part of the image processing was carried out using {\it VST-tube} \citep{Gradoetal2012}, which is a data reduction pipeline specifically developed for the VST-OmegaCAM mosaics. It performs pre-reduction, astrometric and photometric calibration and mosaic production.

To remove the instrumental signatures, after overscan, bias and flat-field corrections, we performed gain equalization of the 32 CCDs and illumination correction. The last step was achieved by comparing the magnitude of stars with the SDSS DR12 catalogue \citep{Alametal2015} in equatorial fields observed during the same nights.  
The magnitude difference as function of the position over the whole field of view is fitted using the ``generalized additive method" \citep{Hastie&Tibshirani1986}. This approach has been shown to be more robust compared to a plain surface polynomial fit in particular for what concerns the behavior at the edges of the image. Then, the position-dependent zero-point correction was applied to science images.  The relative photometric calibration of the images was obtained by minimizing the quadratic sum of differences in magnitude between sources in overlapping observations. The tool used for both the astrometric and relative photometric calibration tasks is {\tt SCAMP} \citep{Bertin2006}. The used reference astrometric catalogue is 2MASS \citep{Skrutskieetal2006}. The absolute photometric calibration was obtained with the {\tt Photcal} tool \citep{Radovichetal2004} using as reference the equatorial photometric standard star fields from SDSS DR12 \citep{Alametal2015}. 
Finally the two dithered images for each pointing were co-added. In order to simplify the subsequent image subtraction analysis, for each pointing the mosaics at the different epochs were registered and aligned to the same pixel grid. This assures that each pixel in the mosaic frame corresponds to the same sky coordinates for all epochs. For further details on the data reduction see \citet{Capacciolietal2015}. 

The detection of transients is affected by the presence of artifacts on the images. These artifacts can be due to the telescope, like spikes and haloes around bright stars, and/or to the camera, like electronic cross-talk. It is desirable to detect and correct them or at least to identify the pixels affected  by the artifact. With this purpose we produce a flag image with specific code for different kind of contaminants that can be used during image analysis.
The procedure to mask the spikes around bright stars is described in appendix A whereas in the following we describe the approach adopted for cross talk. 
In fact, two CCDs of the OMEGACAM camera (ccd65 and ccd66 in ESO names coding; c.f. \citep{Mieske2018}) are affected by electronic cross-talk, i.e. the appearing of spurious sources or holes due to the presence of bright sources in the corresponding pixels of the other CCDs.  This produces spurious sources that can be confused with transients. In principle with the knowledge of the  cross-talk coefficients one can attempt to remove the ghost objects. In practice the coefficients change with time then preventing a safe removal of the spurious  sources. For that reason we decided to adopt a conservative approach masking the pixels affected by cross-talk ghosts with a specific flag value.

With the current available hardware, the time needed to produce a one square degree fully calibrated co-added image is $\sim 20$\,min, including the time for data transfer from Cerro Paranal to Naples and the production of the {\em SExtractor} catalogues.

\subsection{Transient search}

In order to search for variable and transient sources, we applied our analysis procedure first described in \citet{Brocatoetal2018}. The pipeline is based on two independent yet synergistic procedures. One, called {\it ph-pipe}, is based on the comparison of the photometric measurements of all the sources in the VST field obtained at different epochs. The second ({\tt diff-pipe}) is based on the analysis of the difference of images following the approach of the supernova (SN) search program conduced with the VST for several years now \citep{Cappellaroetal2015}. 

 The {\tt ph-pipe} is typically more rapid, providing an excellent quick-look facility able to single out candidates in single epoch campaigns 
 
while the {\tt diff-pipe} is more effective for pinpointing sources projected over extended objects or in case of strong crowding. The two approaches are complementary and since they are independent the comparison of the results allow us to spot possible missed transients.
 
  As discussed in \cite{Brocatoetal2018} the list of candidates contains a large number of spurious objects that can be related to small misalignments of the images, improper flux scaling, incorrect PSF convolution {\it or not} well masked CCD defects and cosmic rays. In most cases spurious transients could be easily identified by visual inspection.  However due to the large number of candidates it would be very time consuming. For this reason both pipelines apply several criteria to attribute a ``score"  to select "bona fide" candidates, based  on their morphological parameters, magnitude variations, positions with respect to nearby galaxies, etc.
 \citep[c.f.][]{Cappellaroetal2015}.
 With this approach we can readily reject most spurious sources.  The remaining candidates go through a further selection after being visually inspected.  
 

In consideration of the largely unknown properties of the possible EM gravitational wave counterpart, we decided not to use model-based priors in the candidate selection.
The main goal of our analysis is therefore to simply identify sources showing a ``significant" brightness variation ( 5$\sigma$ between at least two epochs that in our case corresponds typically to $\sim$ 0.5 magnitudes), either rising or declining in flux during the monitoring. 


 The implementation of the {\tt diff-pipe} requires adequate reference images. Ideally, these images should be obtained before the event to allow for a prompt transient search. If they are not available, as  it is in our case, they can be obtained at later epochs. 
Alternatively, we can use as references archival images obtained with other telescopes. Archival images of the survey area of GW170814 are available from the DECAM survey archive\footnote{http://archive.noao.edu}.  This option is not immune to risk 
due to a possible mismatch of the instrument/telescope band-passes, objects with a peculiar spectrum give rise to spurious transients after image subtraction (notice however that such objects will give a constant residual for all the different VST epochs).

Conservatively, in the case of GW170814 we performed the transient search using as reference the VST late time images whereas DECAM reference images were used to produce the transient light curves (cf. later this section). This was also useful to verify the possible use of archive DECAM images for future prompt searches.

The difference images produced with {\tt hotpants} for the 99 pointings were searched with {\tt Sextractor} resulting in a cumulative list of 3,974,951 variable sources.  
The list is large because we adopted a low detection threshold ($1.5\sigma$  above the background noise) to make sure to get the faintest transients. This also means that most of the detected sources are spurious due to noise fluctuation, image defects, not well removed cosmic rays or imperfect subtraction of bright stars, etc.
Following \cite{Cappellaroetal2015} to filter out most of the spurious candidates we used a scoring algorithm based on several metrics of the detected candidates following the scheme reported in Tab.~\ref{scoretab}.

\begin{table}
\caption{Scoring conditions}\label{scoretab}
\begin{tabular}{lrr}
\hline
condition & score & number \\
\hline
FWHM/avg(FWHM)$ > 2.5$  & $-60$   &  1,415,908  \\
ISOAREA$ < 2\times$avg(FWHM) & $-60$  &     475,738 \\
N$_{\rm pix}$(positive)/N$_{\rm pix}$(Negative) < 0.60 & $-60$  &3,435,173 \\
\\
remaining with score >0 & & 133,956 \\
\\
 2.5 >FWHM/avg(FWHM)$ > 1.75$       & $-30$    &   62,712  \\
2$\times$ avg(FWHM)<ISOAREA<3$\times$avg(FWHM) & $-30$    &   38,138  \\
N$_{\rm pix}$(positive)/N$_{\rm pix}$(Negative) < 0.70 & $-30$ & 68,295 \\
near star mag < 17            & $-30$    & 45,215    \\
near galaxy                   & $+30$    & 6519      \\
\\
remaining with score > 0 &                & 10714 \\ 
\hline
\end{tabular}
\end{table}

The numbers in Tab~\ref{scoretab} includes multiple detections at different {\it epochs}. In fact, the final list with score $>0$ counts 9,342 distinct candidates of which 1,687 with high score ($>=60$). The first condition in Tab~\ref{scoretab} adds a penalty of 60 to objects that appear to be non stellar in the difference image. The second condition helps to remove cosmics that were not detected in the early stages of the image processing. The third condition gives a strong penalty to those sources caused by a local mismatch of the PSF that usually gives a "dipolar" source with positive and negative pixels counts.
Then the scoring is repeated with more stringent parameters but with less heavy penalties. {\it Ratings are applied in order to penalise spurious detections with the exceptions of "near galaxy sources", a condition that raises the rank of these objects to a positive score.}

The scoring algorithm and reference value were chosen after extensive artificial star experiments.
We found that with the above scoring scheme and selecting a score threshold $>0$
we can reject $>99$\% of spurious candidates at the cost of loosing 5\% of the real transients. After adopting a score threshold of 60\% the number of spurious events decreases by a factor $\sim 5$ whereas the fraction of real transients that are lost may increase up to by 15\%.

In this work we visually inspected all candidates with score $>0$ which led to a list of 246 validated transients.

We cross-checked the candidates with public data-sets, e.g. SIMBAD \citep{2000A&AS..143....9W}, NED\footnote{https://ned.ipac.caltech.edu/}, and Skybot \citep{2006ASPC..351..367B}, aiming at identifying known objects.
This allowed us to remove from the list two RR Lyrae,  and 21 asteroids.
We also removed transients that are bona fide variable stars because they coincides with   stellar sources present in all our images and also in the DECAM archive images.

We are left with a total number of 53 candidates (36 were identified by both pipelines and 17 only by {\tt diff-pipe}), see Appendix B.

\begin{figure*}
\begin{center}
\includegraphics[scale=0.75]{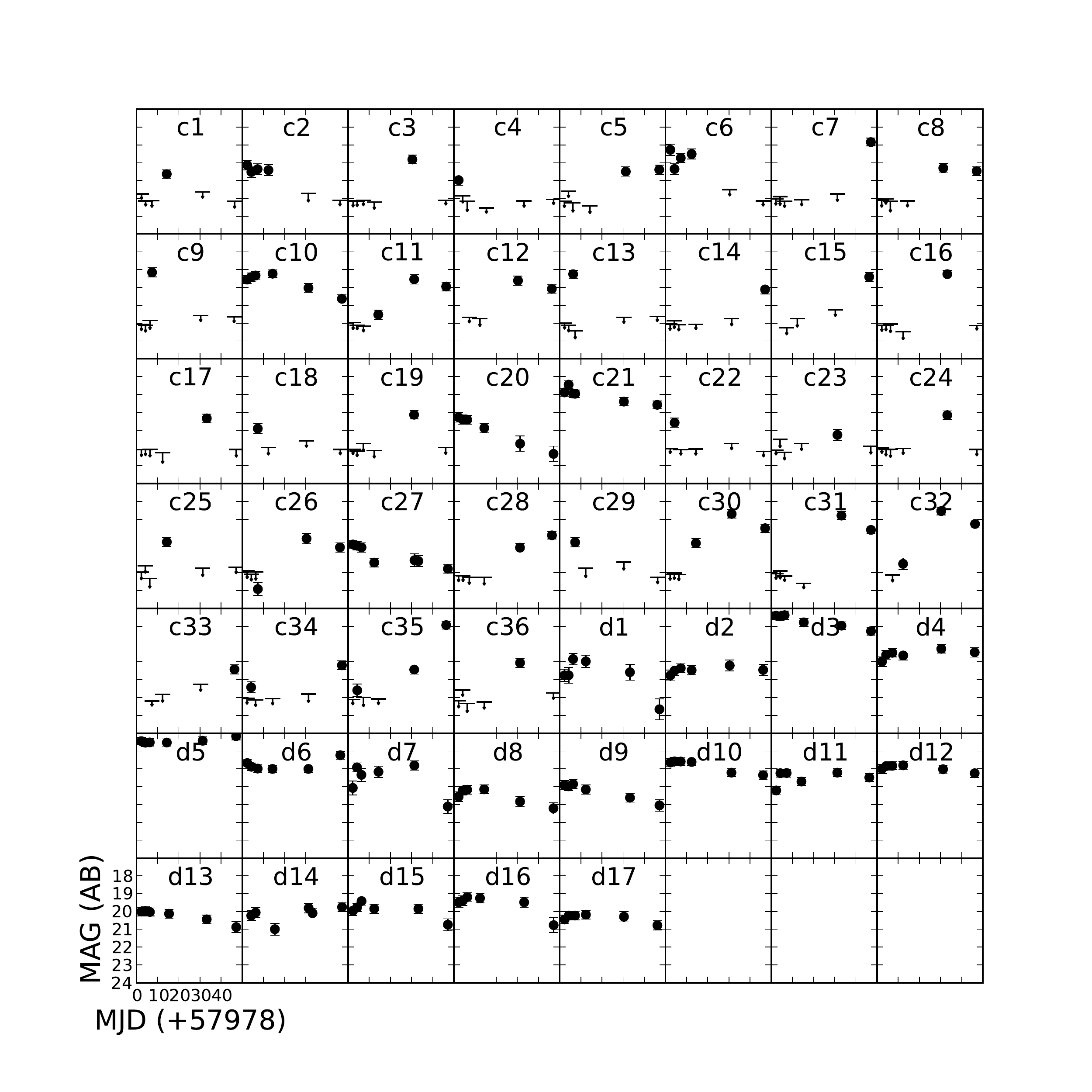}
\end{center}
\caption{ The light curves of the transient candidates found with the VST in $99\,{\rm deg^2}$ covering the GW170814 error area. Arrows indicate the upper limit magnitude for the undetected sources.}
\label{fig:lc}
\end{figure*}

Since  all the selected candidates are also located in the survey area of the Dark Energy Survey, we took DES images observed before the GW event as  reference frames to investigate the sources variability.
Details of the 53 candidates transients, namely their  photometric catalogue  and light curve analysis, are reported in Table  B1  and \ref{tab:transients}. We have differentiate the transients found by both the search pipelines (the prefix {\it c} in the Id) from the ones found only with the image subtraction pipeline (prefix {\it d} in the Id). We reported whether a source was found in the NED (within 5\,arcsec) or SIMBAD (within 3\,arcsec)  (column 4 and 5 of table \ref{tab:transients}, respectively) and the identification that comes from these databases. In figure \ref{fig:lc} we show the light curves for all 53 candidates.  In case of missing detection it is reported the limiting magnitude of the image at the specific epoch.

\begin{table*}
\label{transients}
\centering
\caption{Results from the transient search and identification. In the ID column c indicates that the object was found in both diff-pipe and ph-pipe pipelines, d that the transient was found only with the image subtraction pipeline. In columns 2 and 3 we list RA and DEC (J2000); column 4 and 5 reports  if the source was already present} in NED and SIMBAD database respectively. Column 6, 7 and 8 report average magnitude, magnitude variance and number of available magnitudes including upper limits where available. Column 9 reports the template used and the results of the light curve fit, where z is the best-fit redshift, p is the phase. 
Column 10 lists the probability at the position in the LALinference map for each candidate. The map has been divided in bins of 10\% of integrated probability, the reported number is the  upper limit of the bin. For example the candidate d9 is at a position located in the region of the sky map enclosed between  80 and 90\% of confidence levels. After 90 \% we added two further bins: 90-95\% and 95-99\%. Candidates with values 99 are outside the sky map region for that reason unrelated to the GW event.
In the last column, where applicable, are noted remarks about the identification.
\label{tab:transients}
\tabcolsep=0.09cm
\begin{tabular}{lcccccccccc}
\hline\hline
Id & RA & Dec &  NED & SIMBAD & $mag_{mean}$ & $mag_{var}$ & $N_{mag}$ & light curve fit & Localization &Note\\
\hline
c1 & 47.93876 & -32.50572 & - & - & 20.64 & >2.8 & 1 & - & 99 & 1 \\
c2 & 41.12476 & -47.08170 & - & - & 20.36 & 0.38 & 4 & Ia 1990N z=0.12 p=8 & 80 &  \\
c3  & 44.24308 & -36.10156 &  -  &  -  & 19.82 & >3.1 & 1 &  -  & 99 & 1 \\ 
c4  & 43.55516 & -46.05383 &  -  &  -  & 20.98 & >2.9 & 1 &  AGN   & 80 & \\ 
c5  & 39.56146 & -45.53453 &  -  &  -  & 20.45 & 0.11 & 2 & IIP 1999em	z=0.04 p=15  & 80 & \\ 
c6  & 44.29046 & -37.11974 &  -  &  -  & 19.72 & 1.08 & 4 & Ia faint 1991bg	z=0.06	p=5  & 99 &    \\ 
c7  & 36.37832 & -46.60462 &  Y  &  -  & 18.85 & >3.5 & 1 &  -  & 95 & 1 \\ 
c8  & 35.69231 & -43.94526 &  -  &  -  & 20.39 & 0.17 & 2 & Ia 1994D	z=0.05	p=42  & 99 &   \\ 
c9  & 40.96637 & -39.09182 &  -  &  -  & 19.15 & >2.4 & 1 &  -  & 99 & 1  \\ 
c10  & 42.84391 & -46.75855 &  Y	 & -  & 19.70 & 1.41 & 6 & IcBL 1998bw	z=0.09	p=0	  &  80 & \\ 
c11  & 47.22520 & -33.98899 &  Y	 & -  & 20.34 & 1.98 & 3 & Iapec 2000cx	z=0.08		p=20	  & 95 &  \\ 
c12  & 48.52714 & -42.08370 &  -  &  	-  & 19.85 & 0.47 & 2 &  1990N	z=0.07	p=0	  & 30 &  \\ 
c13  & 40.71350 & -38.97058 &  -  &  -  & 19.25 & >3.7 & 1 &  -  & 99 & 1 \\ 
c14  & 36.19326 & -46.11161 &  -  &  -  & 20.11 & >2.3 & 1 &  -  & 95 &  1 \\ 
c15  & 49.44558 & -43.49652 &  -  &  -  & 19.41 & >3.4 & 1 &  -  & 60 & 1 \\ 
c16  & 43.33967 & -41.94081 &  -  &  -  & 19.25 & >3.6 & 1 &  -  & 95 & 1 \\ 
c17  & 42.54444 & -41.49203 &  -  &  -  & 20.34 & >2.7 & 1 &  -  & 95 &  1 \\ 
c18  & 47.20362 & -41.20692 &  Y  &  -  & 20.91 & >0.6 & 1 &  AGN  & 50 &   \\ 
c19  & 42.06331 & -49.45406 &  -  &  -  & 20.14 & >2.9 & 1 &  -  & 80 &  1 \\ 
c20  & 36.50260 & -43.98644 &  -  &  -  & 21.02 & 2.05 & 6 & Ia 1992A	z=0.07	p=-25  & 95 & \\ 
c21  & 41.17931 & -40.37404 &  	Y	 & - & 19.04 & 1.13 & 6 & Ia 2002bo	z=0.055	p=-10  & 95 &  \\ 
c22  & 36.94201 & -49.98622 &  -  &  -  & 20.59 & >2.1 & 1 &  -  & 80 &  1 \\ 
c23  & 42.92521 & -45.71532 &  -  &  -  & 21.27 & >1.0 & 1 &  -  & 80 &  1 \\ 
c24  & 42.71470 & -42.51475 &  -  &  -  & 20.16 & >3.8 & 1 &  -  & 95 &  1 \\ 
c25  & 40.17324 & -46.87562 &  -  &  -  & 20.27 & >2.9 & 1 &  -  & 70 &  1 \\ 
c26  & 42.22224 & -38.25838 &  -  &  -  & 21.19 & 2.83 & 3 & Ia	1990N	z=0.1	p=22  &  99 & \\ 
c27  & 40.63033 & -41.06565 &  Y  & -	 & 21.05 & 1.37 & 7 & Ia 1994D	z=0.071	p=0	  &  95 & 2  \\ 
c28  & 36.88136 & -52.49284 &  -  &  -  & 20.25 & 0.69 & 2 & Ia	1990N	z=0.095	p=41  &  90 &  \\ 
c29  & 48.50660 & -42.58589 &  -  &  -  & 20.29 & >3.2 & 1 &  -  & 30 &  1 \\ 
c30  & 35.93385 & -44.02682 &  Y	 &  	-  & 19.52 & 1.64 & 3 & SNLC	2008es	z=0.14		p=21  & 99 &   \\ 
c31  & 47.23585 & -46.62792 &  -  &  -  & 19.30 & 0.80 & 3 &  Ia 1992A	z=0.05	p=35  &  20 & \\ 
c32  & 46.59110 & -38.23088 &  -  &  -  & 19.77 & 2.96 & 3 & Ia	1994D	z=0.05	p=20  & 90 & 3 \\ 
c33  & 46.09634 & -42.57915 &  -  &  -  & 20.42 & >1.9 & 1 &  -  &   50 & 1 \\ 
c34  & 38.97934 & -45.03692 &  -  &  -  & 20.81 & 1.24 & 2 &  -  &   80 & 1 \\ 
c35  & 38.86904 & -52.55772 &  -  &  -  & 19.99 & 3.66 & 3 &  -  &   80 & 1 \\ 
c36  & 39.28249 & -45.36645 &  -  &  -  & 20.05 & >2.9 & 1 &  -  &   80 & 1 \\ 
d1  & 42.24100 & -43.27893 &  Y  &  -  & 20.60 & 3.04 & 7 & Ia 1992A z=0.12 p=-6  &  95 & \\ 
d2  & 44.27798 & -36.80744 &  Y	 & -  & 20.45 & 0.56 & 6 & IIP 1999em	z=0.04	p=0	  & 99 &  \\ 
d3  & 46.29475 & -45.55090 &  Y  & 	Galaxy  & 17.76 & 0.90 & 7 &  SLSN	2008es	z=0.0815	p=5  & 99 & 4 \\ 
d4  & 44.32350 & -37.35107 &  Y	 & -  & 19.58 & 0.72 & 6 &  AGN  &  99 & \\ 
d5  & 41.50049 & -46.85482 &  Y	 &  QSO  & 17.44 & 0.36 & 6 & AGN (1.385)  &  80 & 5 \\ 
d6  & 45.05304 & -32.30412 &  -  &  -  & 18.80 & 0.77 & 6 &  AGN   & 99 &  \\ 
d7  & 37.81450 & -46.85080 &  Y	 &  -	 & 19.57 & 2.30 & 6 & Ia 1991bg	z=0.05	p=6	  &  80 & \\ 
d8  & 40.11296 & -46.33456 &  -  &  -  & 20.52 & 1.06 & 6 & Ic 2007gr	z=0.07	p=0	  & 80 &  \\ 
d9  & 44.17079 & -42.58189 &  -  &  -	 & 20.28 & 1.19 & 7 & Ia 1992A	z=0.08	p=2	  & 90 & 6  \\
d10  & 41.56164 & -49.89749 &  Y  &  AGN  & 18.83 & 0.77 & 6 &  	AGN  & 80 &  \\ 
d11  & 47.19114 & -33.94179 &  Y	 &  -  & 19.52 & 0.80 & 6 & 	AGN  &  95 & \\ 
d12  & 39.78474 & -48.51388 &  Y	 & 	-	 & 18.96 & 0.44 & 6 & Ia 1994D	z=0.095	p=0	  &  70 & \\ 
d13  & 45.63721 & -46.35241 &  -  &  -	 & 20.23 & 0.91 & 7 & Ia faint 1991bg	z=0.08		p=2  & 10 &  \\ 
d14  & 45.75856 & -44.83118 &  Y	 &  -  & 20.16 & 1.24 & 6 & Ia 1990N	z=0.1		p=40  & 20 &  \\ 
d15  & 44.42912 & -41.44589 &  Y	 &  -   & 19.85 & 1.41 & 7 & 	AGN  & 90 &  \\ 
d16  & 44.37912 & -42.12675 &  Y		 &  - & 19.50 & 1.83 & 7 & AGN   &  90 & \\ 
d17  & 45.45381 & -35.57484 &  -	 &  - & 20.36 & 0.60 & 6 & 	AGN   &  99 & \\ 
\hline
\end{tabular}
\begin{flushleft}
Notes 
1 -  Peculiar transient or unknown asteroid or possible flare star; 2 - host galaxy redshift from NED; 3 - AT 2017gqz in the Transient Name Server (TNS); 4 - host galaxy redshift from NED, SN2017eni in TNS; 5 - redshift from NED; 6 - AT 2017fat in TNS.
\end{flushleft}
\end{table*}

For transients with only one detection we can exclude that they are normal SNe (of any kind) or AGN (they are not associated with galaxy nuclei). 
We cannot exclude unknown asteroids, peculiar fast transients, flare stars (even if they are much rarer).
When multiple detections for a given transient were available we attempted to perform a transient classification using the approach described in Sect. 4 of \cite{Cappellaroetal2015}. In short, the transient light curve was compared with that of templates SNe of different types assuming as a free parameters the phase and redshift. In practice, for each template we compute a grid of simulated light curves in a range of redshift  by applying the appropriate k-correction and time dilation. The best fit is chosen by chi-square minimization (due to the lack of color information we cannot constrain the extinction that is neglected). In this approach, the shape of the light curve is the main constraint for the template selection while the apparent/absolute magnitude comparison set the redshift for the specific template.
In fact, only in two cases the redshift of the host was known from NED.
Yet, because of the very sparse light curve sampling,  limited number of epochs and lack of color information no firm classification  could be secured. We could only conclude that two dozen transients have the light curve consistent with that of a specific SN type  at a given redshift (and are then considered SN candidates).
In addition, ten events are identified as AGN candidates because they are  hosted in galaxy nucleus and shows erratic variability. {\sl Finally, for nineteen transients, with only one detection 
no specific class could be assigned.}

\begin{figure}
\begin{center}
\includegraphics[scale=0.45]{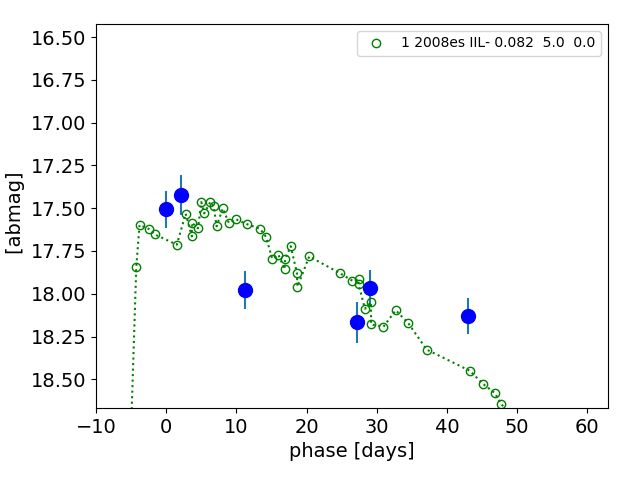}
\includegraphics[scale=0.45]{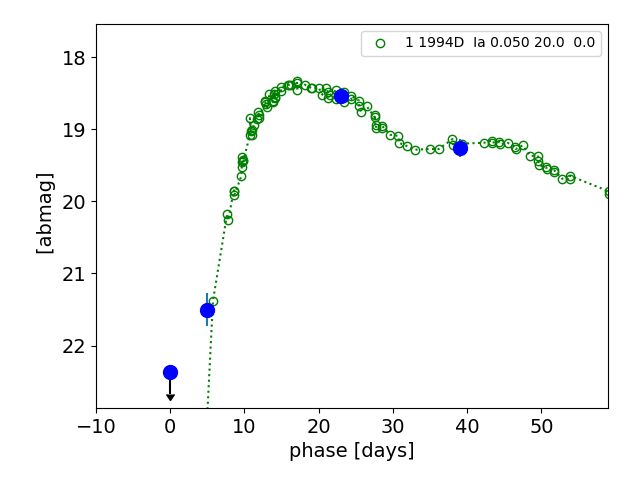}
\includegraphics[scale=0.45]{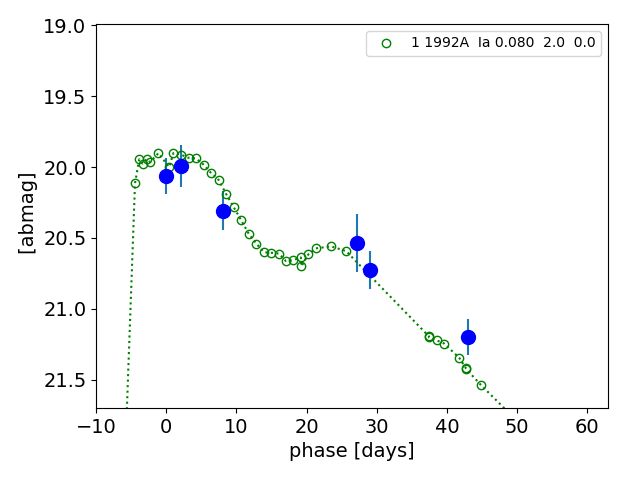}\end{center}
\caption{Top panel: best fit for transient d3 (SN2017eni, blue points) with the template SLSN 2008es (green dots).
Mid panel: best fit for transient c32 (AT 2017gqz, blue points) and the SN Ia 1994D (green dots).
Bottom panel:  best fit for transient d9 (AT 2017fat, blue points) and the SN Ia 1994D (green dots). In the upper right boxes are reported the template used, the redshift and the phase. The last values, always equal to zero, are the extinction.}
\label{fig:fitlc}
\end{figure}

One of the candidate AGN is indeed classified as a QSO in the NED database. Also, three of the SNe candidate are listed in the Transient Name Server (TNS)\footnote{https://wis-tns.weizmann.ac.il/} as follows:

\begin{itemize}
\item 
the transient d3 of table B1 has a light curve shown in Fig. \ref{fig:fitlc} top panel and was obtained by diff-pipe. By assuming the known redshift z=0.0815 \citep{2009ApJ...690.1303M}, we identified it as a super luminous type II SNe similar to SN 2008es.

Indeed we found that the candidate d3 is coincident with SN2017eni (Gaia17blw). This was announced by Gaia Alerts \citep{GaiaAlerts} on 2017 June 6 with G=17.7 mag and indicated as a SN candidate in the galaxy 6dFGS gJ030511.0-453304. It was spectroscopically classified as a Type IIn superluminous supernova by \cite{2017TNSCR.833....1S}.  
Photometry from the ASAS-SN survey finds a peak at V=16.9 mag on 2017 Jun 6.44, implying an absolute magnitude of $M_V$ = -21.0 mag.

\item
Transient c32 is coincident with AT 2017gqz (Gaia17cgz)  announced as a Gaia transient on 2017-09-08 17:47:02 with G=18.79 mag. Our light curve fitting shows that the transient is consistent with a type Ia SN at redshift 0.05 (see Fig. \ref{fig:fitlc} mid panel).

\item
Transient d9 is coincident with  AT 2017fat (Gaia17bqm) announced as a Gaia transient on 2017-06-05 00:41:45 with G=18.18 mag.
The light curve fitting is consistent with that of a Ia SN at a redshift 0.08 (see Fig. \ref{fig:fitlc} bottom panel).
\end{itemize}

We compared our search with the work carried out by \cite{2019ApJ...873L..24D}. The Table \ref{tab:comparison} summarize the comparison of the two follow-up programs in terms of area covered, filter used, depth, number of epochs, time span of the search, total uncertainty enclosed by the surveyed area.

\begin{table*}
\centering
\caption{Summary of the comparison between VST and DECAM observations. For this work the area coverage reported in the table is the maximum area covered in one of the epoch observed (due to telescope operation constraints not always the whole area was covered). The depth for VST is the 50 \% completeness limit for point like sources while in Doctor et al. is reported the 5$\sigma$ magnitude limit for point-like sources.  The total uncertainty covered refers to the refined localization probability region.}
\label{tab:comparison}
\begin{tabular}{lcccccc}
\\
\hline\hline
Survey & Area coverage  & filters & depth &  number of & time span  & total uncertainty   \\
 & (deg$^2$) & & & epochs & (days) & covered (\%) \\
         \hline

This work &  up to 117 & r & 22 & 6 & 45 & 59 \\ 
Doctor et al. & 225 & i & 23 & 8 &12 & 90 \\
\hline
\end{tabular}
\end{table*}

In \cite{2019ApJ...873L..24D} two candidates with coordinates 1) RA= 42.35047 DEC=-40.32632  and 2) RA=47.63365 and DEC=-36.36045 were found. The second object is outside our searching region {\sl while in the position of the first one we have detected a source at magnitude between 22.5 and 23.5 (as reported by \cite{2019ApJ...873L..24D}), which is too faint to be qualified as a transient in our catalogue.}

\section{Detection efficiency}

 To evaluate the survey efficiency we need to know the expected optical light curve of the transient
following a BBH merger. The latter is very uncertain although most authors agree that the optical  emission should be expected much fainter and more rapid declining than the kilonova following a BNS merger.
 To explore the performance of our search we use as reference the most recent models of \cite{2019ApJ...875...49P} that likely encompass the range of possible electromagnetic transient from BBH merger. The different prediction are characterized by:  a) jet energy in the range $10^{46}$ - $10^{49}$ erg; b) jet opening angle in the range 10-40 deg; c) jet Lorentz factor $\Gamma$ (10 or 100).
All models are computed for an ambient number density 0.01 cm$^{-3}$. We immediately verify that a large class of the \cite{2019ApJ...875...49P} models predict optical transients that,
for the given distance of GW170814, are too faint for our instrumentation.  
In particular all models with energy $\leq$ $10^{47}$ erg are too faint regardless of the $\Gamma$ factor and jet opening angle.

We are just marginally sensitive to  models with energy $10^{48}$ erg only if the jet opening angle is small  and our line of sight is along the jet. In the more favorable case of high $\Gamma$ factor, the transient remain above mag 22 for about one day.

We have a measurable efficiency only for $10^{49}$ erg events characterized by high $\Gamma$ factors.

To estimate the chance of detection of a possible optical transient after BBH we used the  simulation of \cite{2019ApJ...875...49P}, as follows.

For each model:

1 - we compute the predicted light curve in apparent magnitude at the distance of the transient (540 Mpc) for all the different viewing angles;

2 - for each pointing and epoch we verify whether the predicted magnitude is brighter that the magnitude limit.  

3 - by considering the probability area covered by the specific transient and the solid angle fraction covered by the viewing solid angle, we compute the chance of detection for the specific pointing.

4- the chance of detection for the full survey is obtained by summing the probability  for all pointings. In the best case of $\Gamma$ = 100 and jet energy $10^{49}$ erg, with a view angle of 40 deg we obtain a survey efficiency of 5.1  per cent. 

A more accurate estimation of the survey efficiency would require a Monte Carlo simulation where  synthetic sources extracted from a population following the \cite{2019ApJ...875...49P} models are added to the images. Then the pipeline procedure to detect the transients is applied to recover the simulated sources. Considering the low efficiency, even for the best case,  this procedure would not change significantly the results.

\section{Discussion and Conclusions}
\label{concl}

We reported on the search for the optical counterpart for the GW event GW170814, by exploiting the capabilities of the VLT Survey Telescope. 
We covered a search area up to 99\,deg$^2$, corresponding to 59 per cent of the credible region, repeated in six epochs distributed over $\sim$ 2 months. A 50 per cent point source completeness at {\it r} $\simeq  22$\,AB mag was reached in most of the epochs. This  threshold corresponds to a luminosity limit of $L_{optical}$ $ \sim$ $1.4^{+0.7}_{-1}$ in unit of $10^{42}$erg/s  after assuming a light flat spectrum and a GW170814 median distance. The errors on the luminosity  mostly derive from the uncertainty on the GW source distance. 
 To estimate the survey efficiency we have considered the recent model for electromagnetic counterpart from binary black holes mergers described in \cite{2019ApJ...875...49P}. In the most favorable case of a very energetic jet emission ($10^{49}$ erg) and high $\Gamma$ factor the expected magnitude of the transient source, at the distance of GW170814, is estimated to remain brighter than mag 22 for  about a day. Therefore, under these specific conditions the corresponding survey efficiency is considerably low, about $5 \%$.
For optical transients  search we developed a pipeline based on two independent analysis algorithms, one based on source extraction and magnitudes comparison between different epochs and the other based on transient identification obtained through image subtraction techniques. 
We present the catalogue of these transients, including, when possible, matches with the SIMBAD within 2 arcsec or NED catalogues and/or possible identification based on light-curve fitting. We identified two dozen candidates SNe, nine AGN candidates, one QSO. Nineteen transients have not been previously catalogued and since they have only one detection cannot be classified.
 For the sake of completeness we have reported in the paper all (53) candidates found. However, if we consider only the ones that fall into the refined localization map, the number  decreases to 40. After removing $d9$,  which was identified as a GAIA transient before the occurrence of the GW event, we are  left with 39 "bona fide" candidates.\\

 Our first two runs were carried out within two days from the gravitational wave event then the  observations became more sparse (this also because we prioritize observations of the kilonova  AT 2017gfo). It is worth noting that the tiles added to improve the coverage of the refined map started from the third epoch and that they cover almost half of the total contained probability. This implies that our observations tend  to constrain models that predict slow-declining light curves like in \cite{2016AAS...22820803B, 2017MNRAS.464..946S}  where a new BBH formation channel inside AGN discs is proposed. In these models the presence of substantial gas densities around the merging stellar mass BHs can produce strong, slow-transient EM counterpart. In this respect our work is complementary to the work of \cite{2019ApJ...873L..24D} that put  constraints on fast fading transients.
Particularly one of the two candidates that they found is outside our search area and the other one is too faint\\

The VST plan for future electromagnetic follow-up of BBH mergers is to have two or more observing epochs within the first couple of days after the alert to search for fast-declining transients and then keep observing for few weeks to put constraints on those models that predict  slower declining light curves. 

\section*{Acknowledgements}
The authors thank the anonymous reviewer for constructive comments on the manuscript. This paper is based on observations made with the ESO/VST. We acknowledge the usage of the OmegaCam and VST Italian GTO time. We also acknowledge INAF financial support of the project "Gravitational Wave Astronomy with the first detections of adLIGO and adVIRGO experiments". This  research  has  made  use
of  public data of the LIGO Scientific Collaboration and the Virgo Collaboration.  LIGO is funded by the U.S. National Science Foundation.  Virgo is funded by the French Centre National
de  Recherche  Scientifique  (CNRS),  the  Italian  Istituto
Nazionale  della  Fisica  Nucleare  (INFN)  and  the  Dutch Nikhef, with contributions by Polish and Hungarian institutes.
S. Covino acknowledges partial funding from Agenzia Spaziale Italiana-Istituto Nazionale di Astrofisica grant I/004/11/3. A.Rossi acknowledges support from Premiale LBT 2013. We acknowledge partial support by the PRIN-INAF 2016 with the project "Towards the SKA and CTA era: discovery, localisation, and physics of transient sources" (P.I. M. Giroletti). \\
This project used public archival data from the Dark Energy Survey (DES) as distributed by the Science Data Archive at NOAO. Funding for the DES Projects has been provided by the DOE and NSF (USA), MISE (Spain), STFC (UK),
HEFCE (UK), NCSA (UIUC), KICP (U. Chicago), CCAPP (Ohio State), MIFPA (Texas A$\&$M), CNPQ, FAPERJ, FINEP
(Brazil), MINECO (Spain), DFG (Germany) and the collaborating institutions in the Dark Energy Survey, which are
Argonne Lab, UC Santa Cruz, University of Cambridge, CIEMAT-Madrid, University of Chicago, University College
London, DES-Brazil Consortium, University of Edinburgh, ETH Zürich, Fermilab, University of Illinois, ICE (IEECCSIC), IFAE Barcelona, Lawrence Berkeley Lab, LMU München and the associated Excellence Cluster Universe,
University of Michigan, NOAO, University of Nottingham, Ohio State University, OzDES Membership Consortium,
University of Pennsylvania, University of Portsmouth, SLAC National Lab, Stanford University, University of Sussex,
and Texas A$\&$M University.
Based on observations at Cerro Tololo Inter-American Observatory, National Optical Astronomy Observatory (NOAO
Prop. ID 2012B-0001; PI: J. Frieman), which is operated by the Association of Universities for Research in Astronomy
(AURA) under a cooperative agreement with the National Science Foundation.
\\
{\it Facility:} {VST ESO programs: 099.D-0191, 099.D-0568, 0100.D-0022}. 


\appendix
\section{Bright stars spikes masking}
\begin{figure}{}
\centering
\includegraphics[width=0.4\textwidth]{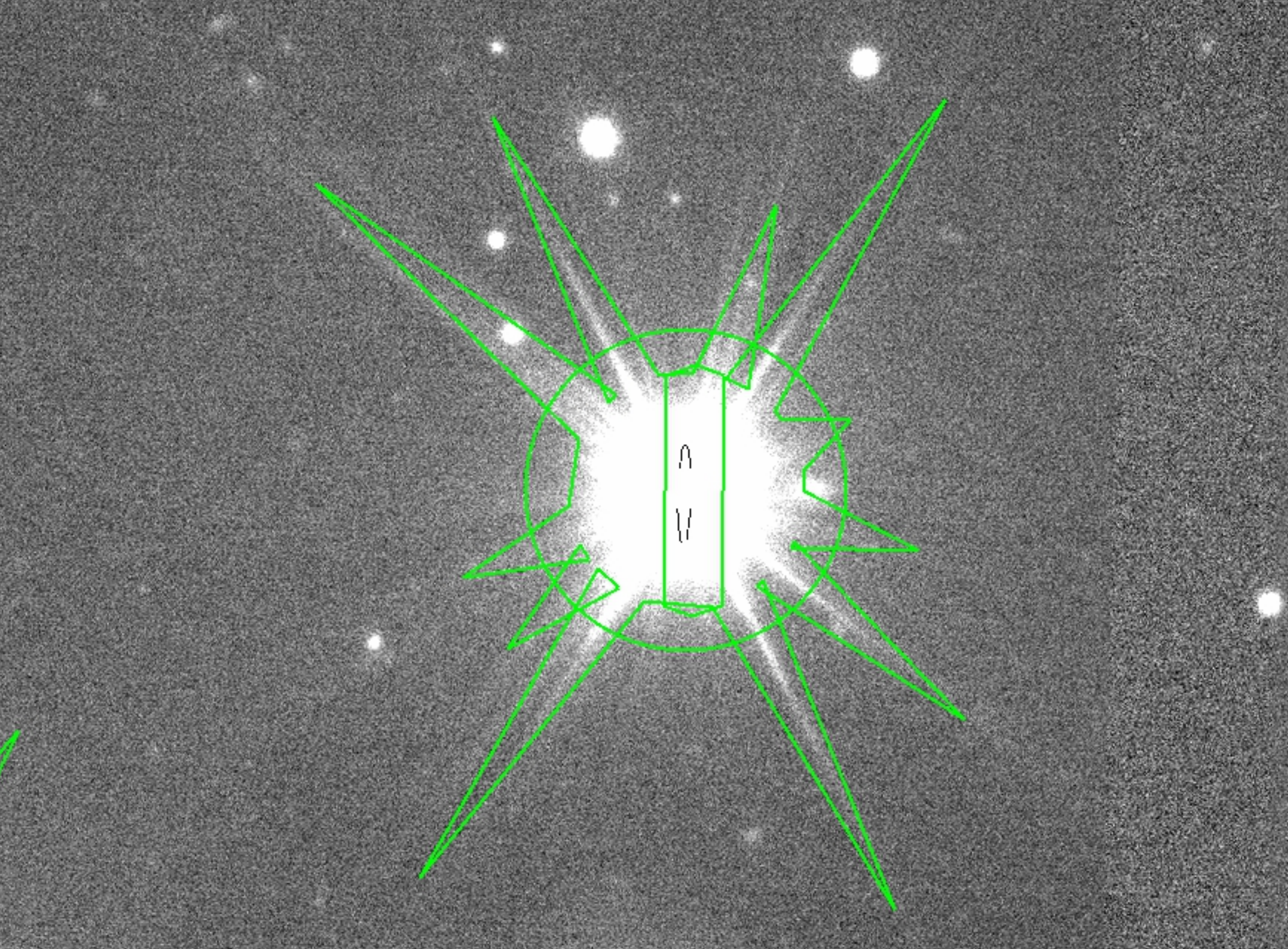}
\caption{Example of bright star mask}
\label{fig:diffmask}
\end{figure}

Here we describe the procedure implemented in VST-Tube to mask haloes and spikes caused by bright stars on OmegaCam images. The spikes are due to diffraction from secondary mirror supports which in case of VST telescope are two pairs of tin supports diametrically opposite to the central secondary mirror. In case of very bright stars, too deep pixels saturation causes an overflow of charges in the pixels along the CCD reading direction. The halos are instead due to multiple reflections at optical surfaces\footnote{http://casu.ast.cam.ac.uk/surveys-projects/vista/technical/spikes-and-halos}. All these artifacts should be masked in order to reduce the contamination of spurious detections in the catalogues extracted from the images. 
The extent to which it is necessary to mask such artifacts is related to the specific analysis performed on the images. In other words the masking should be tailored as much as possible to scientific goals. Our guideline to define the masked regions are the spurious detections (with a SExtractor detection threshold of 1.5 ) on the difference images produced with HOTPANTS \citep{2015ascl.soft04004B}.
For each of the three classes of artifacts mentioned we need to adopt a specific method to create the mask. 
Empirically, we notice that the halo around the stars is a function of the brightness and position of the star with respect to the center of the field. In order to define this function we created a list containing stars position, magnitude and size of the haloes around the stars as determined by visual inspection. The list was used to fit a second order polynomial surface.  For a given exposure time the halo size appears to be quite stable. 
When we need to create the mask for a new image, the first step is to obtain magnitude and position of the bright stars. For that purpose we use an external catalogue \citep[Tycho;][]{Hogetal2000}, accessed through the {\tt astroquery}\footnote{https://astroquery.readthedocs.io/en/latest/} package since the catalogue extracted with SEXtractor is not reliable for very bright stars. The radius of the halo around the bright stars obtained using function previously described, are used also to scale properly the mask of spikes. We can easily recognize that the spikes pattern is quite stable and rotates according to the camera absolute rotation angle (reported in the Fits header). The amplitude of each spike is a function of the star magnitude. We can parametrize the polygonal region that covers the spike with the camera absolute rotation angle and the halo radius (which is a function of the magnitude). Each cusp masking one spike is written as follows:

\begin{equation} 
\label{cusp}
\begin{split}
p_x & =  c_x + (r +d)*sin(\alpha) \\
p_y  & = c_y + (r +d)*cos(\alpha) \\
p_{px}  & =  c_x + r*sin(\alpha + \beta) \\
p_{py}   & = c_y + r*cos(\alpha + \beta) \\
p_{mx}  & = c_x + r*sin(\alpha - \beta) \\
p_{my}  & = c_y + r*cos(\alpha - \beta) \\
\end{split}
\end{equation}

where $c_x$ and $c_y$ are the star coordinates, $r$ is the halo radius, $d$ is the size of the spike being considered, $\alpha$ is the absolute rotation angle and $\beta$ is the angular width at the cusp base. In the present implementations eleven cusps are defined for each star.
Fig. \ref{fig:diffmask} shows an example of the mask for a bright star.

\vspace{3cm}

\section{Transients}

In figures B1 and B2 detailed images  of the 53 transients candidates are shown. Each cut-out, extracted from the original image at the epoch of brightest magnitude, is centered on the transient and has a size of $1'x1'$.
In table B1 is reported  the transients photometry catalogue. In case of missing detection in a particular epoch, we report the limiting magnitude of the image preceded by the symbol {\it >}. 

\clearpage
\begin{figure*}
\begin{center}
\includegraphics[scale=0.25]{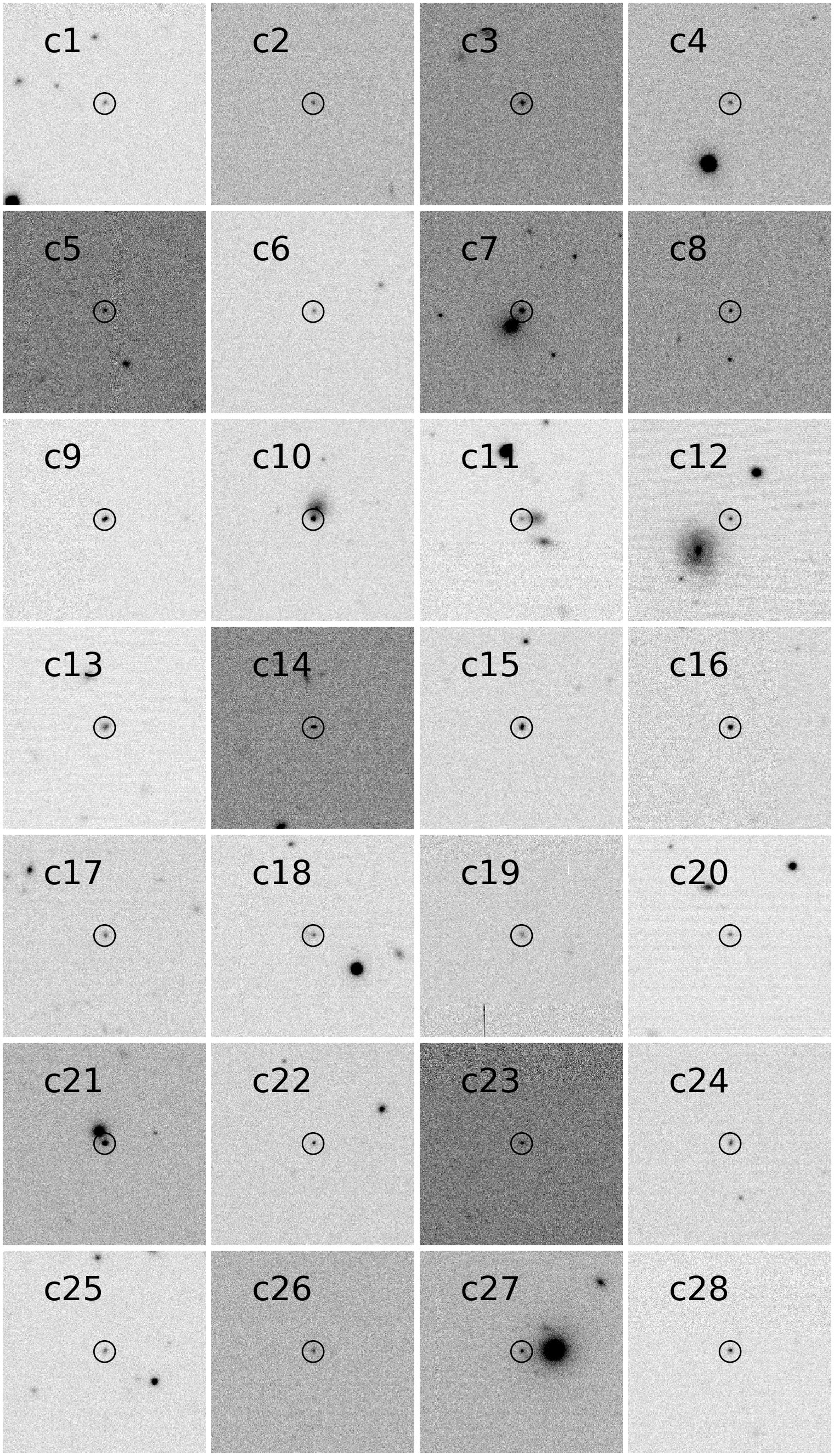}
\end{center}
\label{fig:candidates1}
\caption{First 28 transient candidates found with the VST in $99\,{\rm deg^2}$ covering the GW170814 error area. Each image has a size of $1'x1'$ and is extracted from the original images at the epoch of the brightest magnitude.}
\end{figure*}

\begin{figure*}
\begin{center}
\includegraphics[scale=0.25]{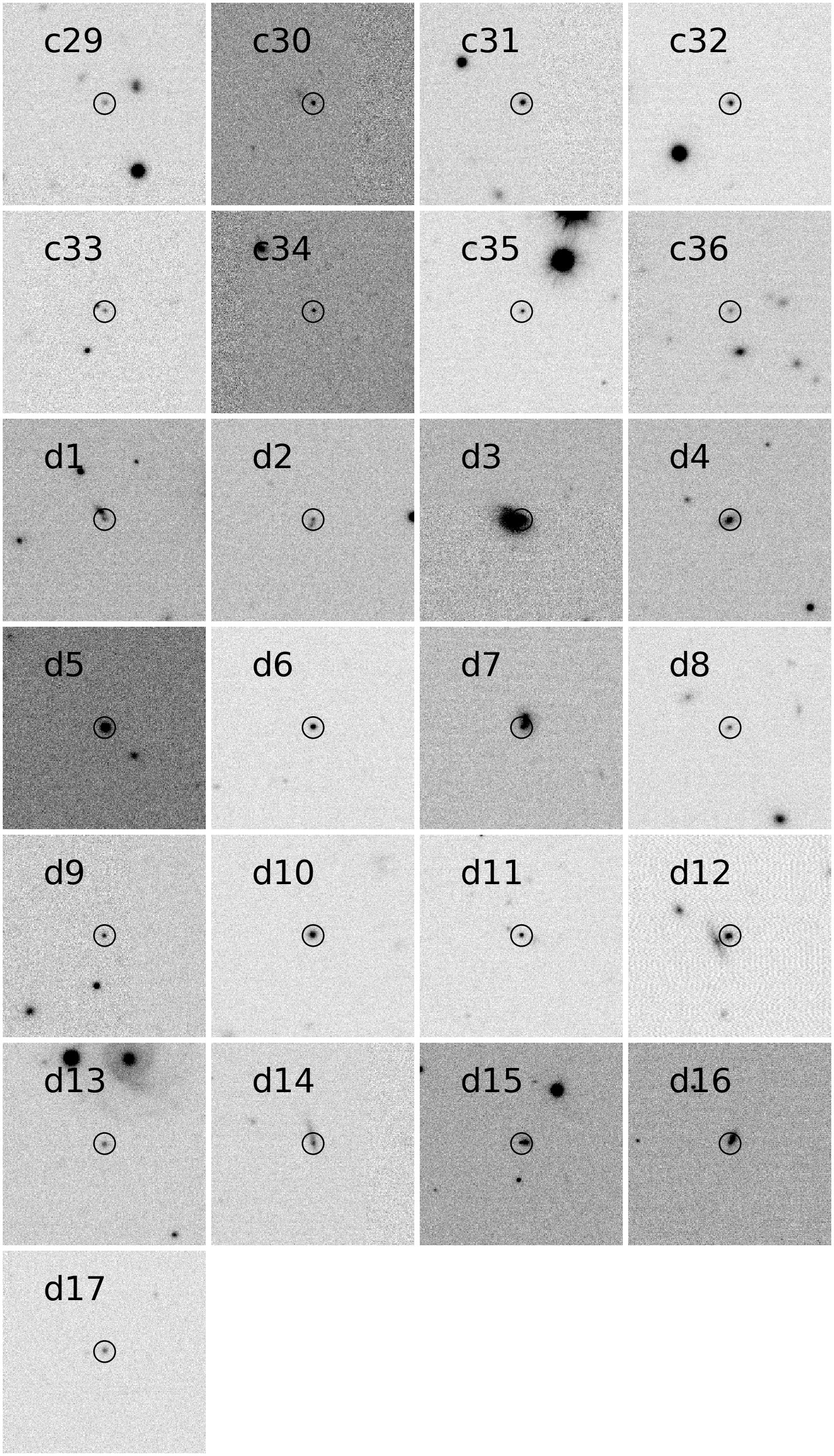}
\end{center}
\label{fig:candidates2}
\caption{Images of transient candidates from c29 to d17 at the brightest magnitude. The size is $1'x1'$.}
\end{figure*}

\begin{table*}
\label{tab:phot_transients}
\centering
\caption{Transients catalogue for the first 3 epochs. Column 1 is the identification name; columns 2 and 3 the coordinates; column 4 is the modified Julian day; column 5 the AB magnitude with the relative error. Then the modified Julian day and magnitude are repeated for the next two epochs. In the epochs without detection it is reported the limiting magnitude preceded by the symbol $>$.}
\label{tab:phottra1}
\tabcolsep=0.11cm
\renewcommand{\arraystretch}{1.1}
\begin{tabular}{lllllllll}
\hline\hline
 &  &  & \multicolumn{2}{|c|}{ epoch 1} & \multicolumn{2}{|c|}{ epoch 2} & \multicolumn{2}{|c|}{ epoch 3} \\
\hline
Id & RA (deg) & DEC (deg) & MJD & MAG (AB) & MJD & MAG (AB) & MJD & MAG (AB)  \\
\hline
c1 & 47.93876 & -32.50572 & 57980.38 & >21.75 $\pm$ 0.20 & 57982.33 & >22.15 $\pm$ 0.18 & 57985.28 & >22.13 $\pm$ 0.27 \\
c2 & 41.12476 & -47.08170 & 57980.34 & 20.14 $\pm$ 0.17 & 57982.29 & 20.52 $\pm$ 0.20 & 57985.24 & 20.36 $\pm$ 0.20 \\
c3 & 44.24308 & -36.10156 & 57980.35 & >22.15 $\pm$ 0.23 & 57982.30 & >22.16 $\pm$ 0.20 & 57985.25 & >22.11 $\pm$ 0.20 \\
c4 & 43.55516 & -46.05383 & 57980.27 & 20.98 $\pm$ 0.19 & 57982.21 & >21.86 $\pm$ 0.30 & 57984.33 & >22.17 $\pm$ 0.47 \\
c5 & 39.56146 & -45.53453 & 57980.26 & >22.16 $\pm$ 0.25 & 57982.18 & >21.60 $\pm$ 0.25 & 57984.29 & >22.25 $\pm$ 0.42 \\
c6 & 44.29046 & -37.11974 & 57980.34 & 19.28 $\pm$ 0.21 & 57982.30 & 20.36 $\pm$ 0.19 & 57985.25 & 19.74 $\pm$ 0.16 \\
c7 & 36.37832 & -46.60462 & 57980.20 & >22.03 $\pm$ 0.25 & 57982.19 & >21.90 $\pm$ 0.39 & 57984.31 & >22.16 $\pm$ 0.25 \\
c8 & 35.69231 & -43.94526 & 57980.21 & >22.10 $\pm$ 0.29 & 57982.20 & >22.03 $\pm$ 0.20 & 57984.32 & >22.16 $\pm$ 0.50 \\
c9 & 40.96637 & -39.09182 & 57980.32 & >22.07 $\pm$ 0.25 & 57982.26 & >22.13 $\pm$ 0.27 & 57985.36 & 19.15 $\pm$ 0.16 \\
c10 & 42.84391 & -46.75855 & 57980.26 & 19.56 $\pm$ 0.12 & 57982.18 & 19.42 $\pm$ 0.13 & 57984.29 & 19.33 $\pm$ 0.11 \\
c11 & 47.22520 & -33.98899 & 57980.37 & >21.97 $\pm$ 0.29 & 57982.33 & >22.11 $\pm$ 0.16 & 57985.28 & >22.16 $\pm$ 0.23 \\
c12 & 48.52714 & -42.08370 & - & - - - & - & - - - & 57985.32 & >21.68 $\pm$ 0.20 \\
c13 & 40.71350 & -38.97058 & 57980.32 & >22.01 $\pm$ 0.22 & 57982.26 & >22.12 $\pm$ 0.27 & 57985.36 & >22.42 $\pm$ 0.36 \\
c14 & 36.19326 & -46.11161 & 57980.20 & >22.05 $\pm$ 0.25 & 57982.19 & >21.87 $\pm$ 0.34 & 57984.31 & >22.09 $\pm$ 0.24 \\
c15 & 49.44558 & -43.49652 & - & - - - & - & - - - & 57985.31 & >22.25 $\pm$ 0.30 \\
c16 & 43.33967 & -41.94081 & 57980.30 & >22.13 $\pm$ 0.23 & 57982.24 & >22.12 $\pm$ 0.21 & 57984.36 & >22.05 $\pm$ 0.38 \\
c17 & 42.54444 & -41.49203 & 57980.31 & >22.10 $\pm$ 0.28 & 57982.25 & >22.10 $\pm$ 0.23 & 57984.37 & >22.07 $\pm$ 0.34 \\
c18 & 47.20362 & -41.20692 & - & - - - & - & - - - & 57985.32 & 20.91 $\pm$ 0.17 \\
c19 & 42.06331 & -49.45406 & 57980.33 & >22.08 $\pm$ 0.19 & 57982.28 & >22.18 $\pm$ 0.21 & 57985.23 & >21.75 $\pm$ 0.34 \\
c20 & 36.50260 & -43.98644 & 57980.22 & 20.28 $\pm$ 0.17 & 57982.21 & 20.41 $\pm$ 0.16 & 57984.32 & 20.42 $\pm$ 0.15 \\
c21 & 41.17931 & -40.37404 & 57980.31 & 18.89 $\pm$ 0.11 & 57982.25 & 18.45 $\pm$ 0.12 & 57985.35 & 18.96 $\pm$ 0.11 \\
c22 & 36.94201 & -49.98622 & 57980.25 & >22.04 $\pm$ 0.17 & 57982.35 & 20.59 $\pm$ 0.16 & 57985.30 & >22.10 $\pm$ 0.20 \\
c23 & 42.92521 & -45.71532 & 57980.26 & >22.13 $\pm$ 0.16 & 57982.18 & >21.52 $\pm$ 0.36 & 57984.29 & >22.25 $\pm$ 0.32 \\
c24 & 42.71470 & -42.51475 & 57980.30 & >22.01 $\pm$ 0.19 & 57982.24 & >22.10 $\pm$ 0.25 & 57984.36 & >22.07 $\pm$ 0.36 \\
c25 & 40.17324 & -46.87562 & 57980.29 & >21.97 $\pm$ 0.32 & 57982.17 & >21.62 $\pm$ 0.30 & 57984.28 & >22.32 $\pm$ 0.44 \\
c26 & 42.22224 & -38.25838 & 57980.33 & >21.90 $\pm$ 0.33 & 57982.27 & >22.09 $\pm$ 0.27 & 57985.36 & 22.92 $\pm$ 0.25 \\
c27 & 40.63033 & -41.06565 & 57980.31 & 20.41 $\pm$ 0.12 & 57982.24 & 20.49 $\pm$ 0.13 & 57984.36 & 20.58 $\pm$ 0.16 \\
c28 & 36.88136 & -52.49284 & 57980.23 & >22.16 $\pm$ 0.24 & 57982.34 & >22.17 $\pm$ 0.22 & 57985.29 & >22.25 $\pm$ 0.30 \\
c29 & 48.50660 & -42.58589 & - & - - - & - & - - - & 57985.32 & 20.29 $\pm$ 0.15 \\
c30 & 35.93385 & -44.02682 & 57980.21 & >22.10 $\pm$ 0.22 & 57982.20 & >22.02 $\pm$ 0.26 & 57984.32 & >22.11 $\pm$ 0.23 \\
c31 & 47.23585 & -46.62792 & 57980.27 & >22.05 $\pm$ 0.22 & 57982.22 & >21.90 $\pm$ 0.35 & 57984.33 & >22.19 $\pm$ 0.20 \\
c32 & 46.59110 & -38.23088 & - & - - - & - & - - - & 57985.34 & >22.12 $\pm$ 0.29 \\
c33 & 46.09634 & -42.57915 & - & - - - & - & - - - & 57985.31 & >22.20 $\pm$ 0.18 \\
c34 & 38.97934 & -45.03692 & 57980.24 & >22.05 $\pm$ 0.22 & 57982.19 & 21.43 $\pm$ 0.20 & 57984.31 & >22.14 $\pm$ 0.26 \\
c35 & 38.86904 & -52.55772 & 57980.24 & >22.11 $\pm$ 0.19 & 57982.34 & 21.59 $\pm$ 0.25 & 57985.29 & >21.99 $\pm$ 0.42 \\
c36 & 39.28249 & -45.36645 & 57980.26 & >22.18 $\pm$ 0.30 & 57982.18 & >21.58 $\pm$ 0.25 & 57984.29 & >22.33 $\pm$ 0.41 \\
d1 & 42.24100 & -43.27893 & 57980.29 & 20.76 $\pm$ 0.24 & 57982.24 & 20.75 $\pm$ 0.34 & 57984.35 & 19.84 $\pm$ 0.21 \\
d2 & 44.27798 & -36.80744 & 57980.35 & 20.76 $\pm$ 0.20 & 57982.30 & 20.50 $\pm$ 0.17 & 57985.25 & 20.35 $\pm$ 0.15 \\
d3 & 46.29475 & -45.55090 & 57980.28 & 17.41 $\pm$ 0.11 & 57982.22 & 17.44 $\pm$ 0.13 & 57984.34 & 17.38 $\pm$ 0.13 \\
d4 & 44.32350 & -37.35107 & 57980.34 & 19.98 $\pm$ 0.17 & 57982.30 & 19.61 $\pm$ 0.15 & 57985.25 & 19.49 $\pm$ 0.14 \\
d5 & 41.50049 & -46.85482 & 57980.29 & 17.45 $\pm$ 0.11 & 57982.17 & 17.53 $\pm$ 0.11 & 57984.28 & 17.52 $\pm$ 0.11 \\
d6 & 45.05304 & -32.30412 & 57980.37 & 18.68 $\pm$ 0.11 & 57982.33 & 18.88 $\pm$ 0.11 & 57985.28 & 18.98 $\pm$ 0.11 \\
d7 & 37.81450 & -46.85080 & 57980.22 & 20.07 $\pm$ 0.29 & 57982.20 & 18.91 $\pm$ 0.15 & 57984.31 & 19.34 $\pm$ 0.27 \\
d8 & 40.11296 & -46.33456 & 57980.25 & 20.56 $\pm$ 0.17 & 57982.17 & 20.21 $\pm$ 0.15 & 57984.28 & 20.17 $\pm$ 0.14 \\
d9 & 44.17079 & -42.58189 & 57980.30 & 19.91 $\pm$ 0.15 & 57982.24 & 19.98 $\pm$ 0.14 & 57984.36 & 19.85 $\pm$ 0.14 \\
d10 & 41.56164 & -49.89749 & 57980.33 & 18.64 $\pm$ 0.12 & 57982.28 & 18.58 $\pm$ 0.12 & 57985.23 & 18.59 $\pm$ 0.12 \\
d11 & 47.19114 & -33.94179 & 57980.37 & 20.21 $\pm$ 0.14 & 57982.33 & 19.24 $\pm$ 0.12 & 57985.28 & 19.23 $\pm$ 0.12 \\
d12 & 39.78474 & -48.51388 & 57980.33 & 19.01 $\pm$ 0.15 & 57982.29 & 18.85 $\pm$ 0.13 & 57985.23 & 18.83 $\pm$ 0.12 \\
d13 & 45.63721 & -46.35241 & 57980.27 & 20.00 $\pm$ 0.14 & 57982.21 & 19.97 $\pm$ 0.14 & 57984.33 & 20.03 $\pm$ 0.13 \\
d14 & 45.75856 & -44.83118 & - & - - - & 57982.23 & 20.23 $\pm$ 0.17 & 57984.34 & 20.06 $\pm$ 0.17 \\
d15 & 44.42912 & -41.44589 & 57980.31 & 19.96 $\pm$ 0.16 & 57982.25 & 19.77 $\pm$ 0.14 & 57984.37 & 19.42 $\pm$ 0.13 \\
d16 & 44.37912 & -42.12675 & 57980.30 & 19.49 $\pm$ 0.18 & 57982.24 & 19.39 $\pm$ 0.17 & 57984.36 & 19.19 $\pm$ 0.14 \\
d17 & 45.45381 & -35.57484 & 57980.36 & 20.45 $\pm$ 0.15 & 57982.31 & 20.23 $\pm$ 0.14 & 57985.26 & 20.23 $\pm$ 0.14 \\
\hline
\end{tabular}
\end{table*}

\begin{table*}
\addtocounter{table}{-1}
\label{tab:phottran2}
\centering
\caption{Transients catalogue for the last 3 epochs. For the description see the caption of the previous table.}
\tabcolsep=0.11cm
\renewcommand{\arraystretch}{1.1}
\begin{tabular}{lllllllll}
\hline\hline
 &  &  & \multicolumn{2}{|c|}{ epoch 4} & \multicolumn{2}{|c|}{ epoch 5} & \multicolumn{2}{|c|}{ epoch 6} \\
\hline
Id & RA (deg) & DEC (deg) & MJD & MAG (AB) & MJD & MAG (AB) & MJD & MAG (AB)  \\
\hline
c1 & 47.93876 & -32.50572 & 57992.29 & 20.64 $\pm$ 0.13 & 58009.24 & >21.64 $\pm$ 0.25 & 58024.36 & >22.17 $\pm$ 0.28 \\
c2 & 41.12476 & -47.08170 & 57990.38 & 20.41 $\pm$ 0.21 & 58009.22 & >21.71 $\pm$ 0.39 & 58024.17 & >22.11 $\pm$ 0.21 \\
c3 & 44.24308 & -36.10156 & 57990.39 & >22.20 $\pm$ 0.31 & 58008.38 & 19.82 $\pm$ 0.15 & 58024.19 & >22.11 $\pm$ 0.16 \\
c4 & 43.55516 & -46.05383 & 57993.38 & >22.54 $\pm$ 0.19 & 58011.19 & >22.15 $\pm$ 0.25 & 58025.14 & >22.06 $\pm$ 0.19 \\
c5 & 39.56146 & -45.53453 & 57992.34 & >22.41 $\pm$ 0.34 & 58009.28 & 20.50 $\pm$ 0.16 & 58025.08 & 20.39 $\pm$ 0.16 \\
c6 & 44.29046 & -37.11974 & 57990.39 & 19.52 $\pm$ 0.18 & 58008.38 & >21.51 $\pm$ 0.25 & 58024.18 & >22.14 $\pm$ 0.19 \\
c7 & 36.37832 & -46.60462 & 57992.38 & >22.07 $\pm$ 0.25 & 58009.30 & >21.75 $\pm$ 0.32 & 58025.10 & 18.85 $\pm$ 0.12 \\
c8 & 35.69231 & -43.94526 & 57992.39 & >22.15 $\pm$ 0.23 & 58009.31 & 20.30 $\pm$ 0.16 & 58025.11 & 20.47 $\pm$ 0.15 \\
c9 & 40.96637 & -39.09182 & - & - - - & 58008.36 & >21.58 $\pm$ 0.23 & 58024.14 & >21.64 $\pm$ 0.23 \\
c10 & 42.84391 & -46.75855 & 57992.34 & 19.23 $\pm$ 0.11 & 58009.28 & 20.03 $\pm$ 0.15 & 58025.08 & 20.63 $\pm$ 0.13 \\
c11 & 47.22520 & -33.98899 & 57992.28 & 21.52 $\pm$ 0.16 & 58009.24 & 19.54 $\pm$ 0.17 & 58024.35 & 19.95 $\pm$ 0.15 \\
c12 & 48.52714 & -42.08370 & 57990.33 & >21.75 $\pm$ 0.33 & 58008.31 & 19.61 $\pm$ 0.16 & 58024.31 & 20.08 $\pm$ 0.14 \\
c13 & 40.71350 & -38.97058 & - & - - - & 58008.36 & >21.68 $\pm$ 0.23 & 58024.14 & >21.63 $\pm$ 0.18 \\
c14 & 36.19326 & -46.11161 & 57992.38 & >22.06 $\pm$ 0.20 & 58009.30 & >21.75 $\pm$ 0.30 & 58025.10 & 20.11 $\pm$ 0.15 \\
c15 & 49.44558 & -43.49652 & 57990.33 & >21.75 $\pm$ 0.37 & 58008.31 & >21.25 $\pm$ 0.27 & 58024.31 & 19.41 $\pm$ 0.14 \\
c16 & 43.33967 & -41.94081 & 57990.36 & >22.48 $\pm$ 0.35 & 58011.23 & 19.25 $\pm$ 0.12 & 58025.17 & >22.14 $\pm$ 0.15 \\
c17 & 42.54444 & -41.49203 & 57990.36 & >22.27 $\pm$ 0.49 & 58011.23 & 20.34 $\pm$ 0.14 & 58025.17 & >22.09 $\pm$ 0.32 \\
c18 & 47.20362 & -41.20692 & 57990.34 & >21.98 $\pm$ 0.31 & 58008.32 & >21.60 $\pm$ 0.26 & 58024.32 & >22.09 $\pm$ 0.19 \\
c19 & 42.06331 & -49.45406 & 57990.37 & >22.15 $\pm$ 0.31 & 58009.20 & 20.14 $\pm$ 0.14 & 58024.16 & >21.98 $\pm$ 0.28 \\
c20 & 36.50260 & -43.98644 & 57992.39 & 20.88 $\pm$ 0.16 & 58009.32 & 21.76 $\pm$ 0.33 & 58025.11 & 22.33 $\pm$ 0.33 \\
c21 & 41.17931 & -40.37404 & - & - - - & 58008.36 & 19.40 $\pm$ 0.13 & 58024.14 & 19.58 $\pm$ 0.12 \\
c22 & 36.94201 & -49.98622 & 57992.37 & >22.06 $\pm$ 0.24 & 58009.27 & >21.75 $\pm$ 0.25 & 58024.38 & >22.19 $\pm$ 0.21 \\
c23 & 42.92521 & -45.71532 & 57992.34 & >21.75 $\pm$ 0.29 & 58009.29 & 21.27 $\pm$ 0.21 & 58025.09 & >21.90 $\pm$ 0.34 \\
c24 & 42.71470 & -42.51475 & 57990.35 & >22.02 $\pm$ 0.25 & 58011.22 & 20.16 $\pm$ 0.14 & 58025.17 & >22.09 $\pm$ 0.25 \\
c25 & 40.17324 & -46.87562 & 57992.33 & 20.27 $\pm$ 0.15 & 58009.28 & >21.75 $\pm$ 0.36 & 58025.08 & >21.70 $\pm$ 0.25 \\
c26 & 42.22224 & -38.25838 & - & - - - & 58008.37 & 20.09 $\pm$ 0.19 & 58024.15 & 20.57 $\pm$ 0.16 \\
c27 & 40.63033 & -41.06565 & 57990.36 & 21.43 $\pm$ 0.14 & 58011.23 & 21.34 $\pm$ 0.20 & 58025.17 & 21.78 $\pm$ 0.14 \\
c28 & 36.88136 & -52.49284 & 57992.36 & >22.25 $\pm$ 0.34 & 58009.25 & 20.59 $\pm$ 0.14 & 58024.37 & 19.90 $\pm$ 0.11 \\
c29 & 48.50660 & -42.58589 & 57990.33 & >21.75 $\pm$ 0.42 & 58008.31 & >21.41 $\pm$ 0.34 & 58024.31 & >22.25 $\pm$ 0.26 \\
c30 & 35.93385 & -44.02682 & 57992.39 & 20.34 $\pm$ 0.16 & 58009.31 & 18.70 $\pm$ 0.13 & 58025.11 & 19.50 $\pm$ 0.13 \\
c31 & 47.23585 & -46.62792 & 57993.38 & >22.59 $\pm$ 0.21 & 58011.19 & 18.79 $\pm$ 0.12 & 58025.14 & 19.60 $\pm$ 0.11 \\
c32 & 46.59110 & -38.23088 & 57990.31 & 21.50 $\pm$ 0.23 & 58008.34 & 18.54 $\pm$ 0.12 & 58024.33 & 19.26 $\pm$ 0.12 \\
c33 & 46.09634 & -42.57915 & 57990.33 & >21.82 $\pm$ 0.34 & 58008.31 & >21.25 $\pm$ 0.30 & 58024.31 & 20.42 $\pm$ 0.16 \\
c34 & 38.97934 & -45.03692 & 57992.36 & >22.06 $\pm$ 0.25 & 58009.29 & >21.81 $\pm$ 0.36 & 58025.10 & 20.19 $\pm$ 0.14 \\
c35 & 38.86904 & -52.55772 & 57992.36 & >22.08 $\pm$ 0.21 & 58009.26 & 20.43 $\pm$ 0.16 & 58024.37 & 17.94 $\pm$ 0.13 \\
c36 & 39.28249 & -45.36645 & 57992.34 & >22.25 $\pm$ 0.30 & 58009.28 & 20.05 $\pm$ 0.15 & 58025.08 & >21.75 $\pm$ 0.25 \\
d1 & 42.24100 & -43.27893 & 57990.35 & 19.97 $\pm$ 0.24 & 58011.21 & 20.58 $\pm$ 0.35 & 58025.16 & 22.66 $\pm$ 0.49 \\
d2 & 44.27798 & -36.80744 & 57990.39 & 20.46 $\pm$ 0.16 & 58008.38 & 20.20 $\pm$ 0.20 & 58024.19 & 20.45 $\pm$ 0.21 \\
d3 & 46.29475 & -45.55090 & 57993.39 & 17.78 $\pm$ 0.12 & 58011.20 & 17.97 $\pm$ 0.12 & 58025.15 & 18.28 $\pm$ 0.12 \\
d4 & 44.32350 & -37.35107 & 57990.39 & 19.64 $\pm$ 0.15 & 58008.38 & 19.27 $\pm$ 0.14 & 58024.18 & 19.46 $\pm$ 0.17 \\
d5 & 41.50049 & -46.85482 & 57992.33 & 17.52 $\pm$ 0.11 & 58009.28 & 17.43 $\pm$ 0.11 & 58025.08 & 17.17 $\pm$ 0.11 \\
d6 & 45.05304 & -32.30412 & 57992.28 & 19.00 $\pm$ 0.11 & 58009.24 & 19.00 $\pm$ 0.12 & 58024.35 & 18.24 $\pm$ 0.11 \\
d7 & 37.81450 & -46.85080 & 57992.38 & 19.16 $\pm$ 0.22 & 58009.31 & 18.81 $\pm$ 0.15 & 58025.10 & 21.11 $\pm$ 0.28 \\
d8 & 40.11296 & -46.33456 & 57992.33 & 20.15 $\pm$ 0.15 & 58009.28 & 20.84 $\pm$ 0.19 & 58025.08 & 21.21 $\pm$ 0.22 \\
d9 & 44.17079 & -42.58189 & 57990.36 & 20.16 $\pm$ 0.15 & 58011.23 & 20.62 $\pm$ 0.15 & 58025.17 & 21.05 $\pm$ 0.22 \\
d10 & 41.56164 & -49.89749 & 57990.37 & 18.61 $\pm$ 0.12 & 58009.20 & 19.21 $\pm$ 0.12 & 58024.16 & 19.35 $\pm$ 0.13 \\
d11 & 47.19114 & -33.94179 & 57992.28 & 19.71 $\pm$ 0.12 & 58009.24 & 19.21 $\pm$ 0.12 & 58024.35 & 19.49 $\pm$ 0.12 \\
d12 & 39.78474 & -48.51388 & 57990.38 & 18.80 $\pm$ 0.13 & 58009.21 & 19.02 $\pm$ 0.12 & 58024.17 & 19.25 $\pm$ 0.14 \\
d13 & 45.63721 & -46.35241 & 57993.38 & 20.13 $\pm$ 0.14 & 58011.19 & 20.44 $\pm$ 0.14 & 58025.14 & 20.88 $\pm$ 0.20 \\
d14 & 45.75856 & -44.83118 & 57993.40 & 20.99 $\pm$ 0.23 & 58011.20 & 20.09 $\pm$ 0.17 & 58025.15 & 19.75 $\pm$ 0.15 \\
d15 & 44.42912 & -41.44589 & 57990.36 & 19.85 $\pm$ 0.16 & 58011.23 & 19.85 $\pm$ 0.15 & 58025.17 & 20.74 $\pm$ 0.23 \\
d16 & 44.37912 & -42.12675 & 57990.36 & 19.26 $\pm$ 0.15 & 58011.23 & 19.49 $\pm$ 0.18 & 58025.17 & 20.77 $\pm$ 0.32 \\
d17 & 45.45381 & -35.57484 & 57990.40 & 20.18 $\pm$ 0.14 & 58008.39 & 20.29 $\pm$ 0.19 & 58024.20 & 20.77 $\pm$ 0.16 \\
\hline
\end{tabular}
\end{table*}
\FloatBarrier

\onecolumn
\section{Details on the pointings}
In the Table \ref{tab:mosaic_details} we report details on the observations for each pointing i.e. input images names, as reported in the ESO archive,  coordinates center, modified Julian day at the start time of the first image, percentage of the masked area due to bright stars, bad columns and gaps among ccds not covered by the dithering, skymap enclosed probability and $50\% $ completeness for point-like sources.

\setlength{\tabcolsep}{2pt}
\setlength{\LTleft}{1pt}
\setlength{\LTcapwidth}{15cm}
\begin{longtable}{cccccccc}
\label{tab:mosaic_details} \\
\caption{Details on observations for each pointing. \hspace{0.4\textwidth}
}\\

\hline
 Image 1 & Image 2 & Ra & Dec & MJD & Masked & Encl. & Compl. \\
\hline
 & & (deg) & (deg) &  & (\%) & prob. (\%) & (AB)  \\
 \hline
OMEGA.2017-08-15T05:07:05.219 & OMEGA.2017-08-15T05:08:30.230 & 35.4016 & -44.3820 & 57980.2133 & 7.73 & 0.00004 & 22.16 \\ 
OMEGA.2017-08-15T04:58:38.518 & OMEGA.2017-08-15T05:00:03.138 & 35.3759 & -45.4200 & 57980.2074 & 3.29 & 0.00007 & 22.07 \\ 
OMEGA.2017-08-15T04:50:12.206 & OMEGA.2017-08-15T04:51:36.286 & 35.3496 & -46.4579 & 57980.2015 & 3.04 & 0.00005 & 22.02 \\ 
OMEGA.2017-08-15T05:41:45.265 & OMEGA.2017-08-15T05:43:09.485 & 35.7322 & -51.4176 & 57980.2373 & 5.75 & 0.00010 & 22.04 \\
\end{longtable}
\twocolumn

\bibliographystyle{mnras}
\bibliography{gw170814}

\begin{thebibliography}{}
\makeatletter
\relax
\def\mn@urlcharsother{\let\do\@makeother \do\$\do\&\do\#\do\^\do\_\do\%\do\~}
\def\mn@doi{\begingroup\mn@urlcharsother \@ifnextchar [ {\mn@doi@}
  {\mn@doi@[]}}
\def\mn@doi@[#1]#2{\def\@tempa{#1}\ifx\@tempa\@empty \href
  {http://dx.doi.org/#2} {doi:#2}\else \href {http://dx.doi.org/#2} {#1}\fi
  \endgroup}
\def\mn@eprint#1#2{\mn@eprint@#1:#2::\@nil}
\def\mn@eprint@arXiv#1{\href {http://arxiv.org/abs/#1} {{\tt arXiv:#1}}}
\def\mn@eprint@dblp#1{\href {http://dblp.uni-trier.de/rec/bibtex/#1.xml}
  {dblp:#1}}
\def\mn@eprint@#1:#2:#3:#4\@nil{\def\@tempa {#1}\def\@tempb {#2}\def\@tempc
  {#3}\ifx \@tempc \@empty \let \@tempc \@tempb \let \@tempb \@tempa \fi \ifx
  \@tempb \@empty \def\@tempb {arXiv}\fi \@ifundefined
  {mn@eprint@\@tempb}{\@tempb:\@tempc}{\expandafter \expandafter \csname
  mn@eprint@\@tempb\endcsname \expandafter{\@tempc}}}

\bibitem[\protect\citeauthoryear{{Aasi} et~al.,}{{Aasi}
  et~al.}{2015}]{Aasietal2015}
{Aasi} J.,  et~al., 2015, \mn@doi [Classical and Quantum Gravity]
  {10.1088/0264-9381/32/11/115012}, \href
  {http://cdsads.u-strasbg.fr/abs/2015CQGra..32k5012A} {32, 115012}

\bibitem[\protect\citeauthoryear{{Abbott} et~al.,}{{Abbott}
  et~al.}{2017a}]{Abbottetal2017a}
{Abbott} B.~P.,  et~al., 2017a, \mn@doi [Physical Review Letters]
  {10.1103/PhysRevLett.119.141101}, \href
  {http://cdsads.u-strasbg.fr/abs/2017PhRvL.119n1101A} {119, 141101}

\bibitem[\protect\citeauthoryear{{Abbott} et~al.,}{{Abbott}
  et~al.}{2017b}]{Abbottetal2017b}
{Abbott} B.~P.,  et~al., 2017b, \mn@doi [Physical Review Letters]
  {10.1103/PhysRevLett.119.161101}, \href
  {http://cdsads.u-strasbg.fr/abs/2017PhRvL.119p1101A} {119, 161101}

\bibitem[\protect\citeauthoryear{{Acernese} et~al.,}{{Acernese}
  et~al.}{2015}]{Acerneseetal2015}
{Acernese} F.,  et~al., 2015, \mn@doi [Classical and Quantum Gravity]
  {10.1088/0264-9381/32/2/024001}, \href
  {http://cdsads.u-strasbg.fr/abs/2015CQGra..32b4001A} {32, 024001}

\bibitem[\protect\citeauthoryear{{Alam} et~al.,}{{Alam}
  et~al.}{2015}]{Alametal2015}
{Alam} S.,  et~al., 2015, \mn@doi [\apjs] {10.1088/0067-0049/219/1/12}, \href
  {http://cdsads.u-strasbg.fr/abs/2015ApJS..219...12A} {219, 12}

\bibitem[\protect\citeauthoryear{{Bartos}}{{Bartos}}{2016}]{2016AAS...22820803B}
{Bartos} I.,  2016, in American Astronomical Society Meeting Abstracts \#228.
  p. 208.03

\bibitem[\protect\citeauthoryear{{Becker}}{{Becker}}{2015}]{2015ascl.soft04004B}
{Becker} A.,  2015, {HOTPANTS: High Order Transform of PSF ANd Template
  Subtraction}, Astrophysics Source Code Library (\mn@eprint {ascl} {1504.004})

\bibitem[\protect\citeauthoryear{{Berthier}, {Vachier}, {Thuillot}, {Fernique},
  {Ochsenbein}, {Genova}, {Lainey}  \& {Arlot}}{{Berthier}
  et~al.}{2006}]{2006ASPC..351..367B}
{Berthier} J.,  {Vachier} F.,  {Thuillot} W.,  {Fernique} P.,  {Ochsenbein} F.,
   {Genova} F.,  {Lainey} V.,   {Arlot} J.-E.,  2006, in {Gabriel} C.,
  {Arviset} C.,  {Ponz} D.,   {Enrique} S.,  eds,  Astronomical Society of the
  Pacific Conference Series Vol. 351, Astronomical Data Analysis Software and
  Systems XV. p.~367

\bibitem[\protect\citeauthoryear{{Bertin}}{{Bertin}}{2006}]{Bertin2006}
{Bertin} E.,  2006, in {Gabriel} C.,  {Arviset} C.,  {Ponz} D.,   {Enrique} S.,
   eds,  Astronomical Society of the Pacific Conference Series Vol. 351,
  Astronomical Data Analysis Software and Systems XV. p.~112

\bibitem[\protect\citeauthoryear{{Bertin}}{{Bertin}}{2009}]{Bertin2009}
{Bertin} E.,  2009, \memsai, \href
  {http://cdsads.u-strasbg.fr/abs/2009MmSAI..80..422B} {80, 422}

\bibitem[\protect\citeauthoryear{{Brocato} et~al.,}{{Brocato}
  et~al.}{2018}]{Brocatoetal2018}
{Brocato} E.,  et~al., 2018, \mn@doi [\mnras] {10.1093/mnras/stx2730}, \href
  {http://cdsads.u-strasbg.fr/abs/2018MNRAS.474..411B} {474, 411}

\bibitem[\protect\citeauthoryear{{Capaccioli}, {Mancini}  \&
  {Sedmak}}{{Capaccioli} et~al.}{2003}]{Capacciolietal2003}
{Capaccioli} M.,  {Mancini} D.,   {Sedmak} G.,  2003, \memsai, \href
  {http://cdsads.u-strasbg.fr/abs/2003MmSAI..74..450C} {74, 450}

\bibitem[\protect\citeauthoryear{{Capaccioli} et~al.,}{{Capaccioli}
  et~al.}{2015}]{Capacciolietal2015}
{Capaccioli} M.,  et~al., 2015, \mn@doi [\aap] {10.1051/0004-6361/201526252},
  \href {http://cdsads.u-strasbg.fr/abs/2015A%26A...581A..10C} {581, A10}

\bibitem[\protect\citeauthoryear{{Cappellaro} et~al.,}{{Cappellaro}
  et~al.}{2015}]{Cappellaroetal2015}
{Cappellaro} E.,  et~al., 2015, \mn@doi [\aap] {10.1051/0004-6361/201526712},
  \href {http://cdsads.u-strasbg.fr/abs/2015A%26A...584A..62C} {584, A62}

\bibitem[\protect\citeauthoryear{{D{\'a}lya} et~al.,}{{D{\'a}lya}
  et~al.}{2018}]{Dalyaetal2018}
{D{\'a}lya} G.,  et~al., 2018, \mn@doi [\mnras] {10.1093/mnras/sty1703}, \href
  {http://cdsads.u-strasbg.fr/abs/2018MNRAS.479.2374D} {479, 2374}

\bibitem[\protect\citeauthoryear{{Delgado}, {Harrison}, {Hodgkin}, {Leeuwen},
  {Rixon}  \& {Yoldas}}{{Delgado} et~al.}{2017}]{GaiaAlerts}
{Delgado} A.,  {Harrison} D.,  {Hodgkin} S.,  {Leeuwen} M.~V.,  {Rixon} G.,
  {Yoldas} A.,  2017, Transient Name Server Discovery Report, \href
  {http://adsabs.harvard.edu/abs/2017TNSTR.632....1D} {632}

\bibitem[\protect\citeauthoryear{{Doctor} et~al.,}{{Doctor}
  et~al.}{2019}]{2019ApJ...873L..24D}
{Doctor} Z.,  et~al., 2019, \mn@doi [\apjl] {10.3847/2041-8213/ab08a3}, \href
  {http://cdsads.u-strasbg.fr/abs/2019ApJ...873L..24D} {873, L24}

\bibitem[\protect\citeauthoryear{{Doi} et~al.,}{{Doi}
  et~al.}{2010}]{Doietal2010}
{Doi} M.,  et~al., 2010, \mn@doi [\aj] {10.1088/0004-6256/139/4/1628}, \href
  {http://adsabs.harvard.edu/abs/2010AJ....139.1628D} {139, 1628}

\bibitem[\protect\citeauthoryear{{Grado}, {Capaccioli}, {Limatola}  \&
  {Getman}}{{Grado} et~al.}{2012}]{Gradoetal2012}
{Grado} A.,  {Capaccioli} M.,  {Limatola} L.,   {Getman} F.,  2012, Memorie
  della Societa Astronomica Italiana Supplementi, \href
  {http://cdsads.u-strasbg.fr/abs/2012MSAIS..19..362G} {19, 362}

\bibitem[\protect\citeauthoryear{{Greco}, {Grado}, {Getman}, {Cappellaro},
  {Branchesi}, {Covino}  \& {GRAWITA-GRavitational Wave Inaf TeAm}}{{Greco}
  et~al.}{2017}]{Grecoetal2017}
{Greco} G.,  {Grado} A.,  {Getman} F.,  {Cappellaro} E.,  {Branchesi} M.,
  {Covino} S.,   {GRAWITA-GRavitational Wave Inaf TeAm} 2017, GRB Coordinates
  Network, Circular Service, No.~21498, \#1 (2017), 21498

\bibitem[\protect\citeauthoryear{Hastie \& Tibshirani}{Hastie \&
  Tibshirani}{1986}]{Hastie&Tibshirani1986}
Hastie T.,  Tibshirani R.,  1986, \mn@doi [Statist. Sci.]
  {10.1214/ss/1177013604}, 1, 297

\bibitem[\protect\citeauthoryear{{H{\o}g} et~al.,}{{H{\o}g}
  et~al.}{2000}]{Hogetal2000}
{H{\o}g} E.,  et~al., 2000, \aap, \href
  {http://adsabs.harvard.edu/abs/2000A%26A...355L..27H} {355, L27}

\bibitem[\protect\citeauthoryear{{Kalogera}, {Belczynski}, {Kim},
  {O'Shaughnessy}  \& {Willems}}{{Kalogera} et~al.}{2007}]{2007PhR...442...75K}
{Kalogera} V.,  {Belczynski} K.,  {Kim} C.,  {O'Shaughnessy} R.,   {Willems}
  B.,  2007, \mn@doi [\physrep] {10.1016/j.physrep.2007.02.008}, \href
  {https://ui.adsabs.harvard.edu/abs/2007PhR...442...75K} {442, 75}

\bibitem[\protect\citeauthoryear{{Kuijken}}{{Kuijken}}{2011}]{Kuijken2011}
{Kuijken} K.,  2011, The Messenger, \href
  {http://cdsads.u-strasbg.fr/abs/2011Msngr.146....8K} {146, 8}

\bibitem[\protect\citeauthoryear{{Mieske}}{{Mieske}}{2018}]{Mieske2018}
{Mieske} S.,  2018, in VST in the Era of the Large Sky Surveys. p.~5,
  \mn@doi{10.5281/zenodo.1303284}

\bibitem[\protect\citeauthoryear{{Miller} et~al.,}{{Miller}
  et~al.}{2009}]{2009ApJ...690.1303M}
{Miller} A.~A.,  et~al., 2009, \mn@doi [\apj] {10.1088/0004-637X/690/2/1303},
  \href {http://cdsads.u-strasbg.fr/abs/2009ApJ...690.1303M} {690, 1303}

\bibitem[\protect\citeauthoryear{{Murase}, {Kashiyama}, {M{\'e}sz{\'a}ros},
  {Shoemaker}  \& {Senno}}{{Murase} et~al.}{2016}]{2016ApJ...822L...9M}
{Murase} K.,  {Kashiyama} K.,  {M{\'e}sz{\'a}ros} P.,  {Shoemaker} I.,
  {Senno} N.,  2016, \mn@doi [\apjl] {10.3847/2041-8205/822/1/L9}, \href
  {http://adsabs.harvard.edu/abs/2016ApJ...822L...9M} {822, L9}

\bibitem[\protect\citeauthoryear{{Pankow}, {Sampson}, {Perri}, {Chase},
  {Coughlin}, {Zevin}  \& {Kalogera}}{{Pankow}
  et~al.}{2017}]{2017ApJ...834..154P}
{Pankow} C.,  {Sampson} L.,  {Perri} L.,  {Chase} E.,  {Coughlin} S.,  {Zevin}
  M.,   {Kalogera} V.,  2017, \mn@doi [\apj] {10.3847/1538-4357/834/2/154},
  \href {https://ui.adsabs.harvard.edu/abs/2017ApJ...834..154P} {834, 154}

\bibitem[\protect\citeauthoryear{{Perna}, {Lazzati}  \& {Giacomazzo}}{{Perna}
  et~al.}{2016}]{2016ApJ...821L..18P}
{Perna} R.,  {Lazzati} D.,   {Giacomazzo} B.,  2016, \mn@doi [\apjl]
  {10.3847/2041-8205/821/1/L18}, \href
  {http://adsabs.harvard.edu/abs/2016ApJ...821L..18P} {821, L18}

\bibitem[\protect\citeauthoryear{{Perna}, {Chruslinska}, {Corsi}  \&
  {Belczynski}}{{Perna} et~al.}{2018}]{2018MNRAS.477.4228P}
{Perna} R.,  {Chruslinska} M.,  {Corsi} A.,   {Belczynski} K.,  2018, \mn@doi
  [\mnras] {10.1093/mnras/sty814}, \href
  {http://adsabs.harvard.edu/abs/2018MNRAS.477.4228P} {477, 4228}

\bibitem[\protect\citeauthoryear{{Perna}, {Lazzati}  \& {Farr}}{{Perna}
  et~al.}{2019}]{2019ApJ...875...49P}
{Perna} R.,  {Lazzati} D.,   {Farr} W.,  2019, \mn@doi [\apj]
  {10.3847/1538-4357/ab107b}, \href
  {https://ui.adsabs.harvard.edu/abs/2019ApJ...875...49P} {875, 49}

\bibitem[\protect\citeauthoryear{{Pian} et~al.,}{{Pian}
  et~al.}{2017}]{2017Natur.551...67P}
{Pian} E.,  et~al., 2017, \mn@doi [\nat] {10.1038/nature24298}, \href
  {https://ui.adsabs.harvard.edu/abs/2017Natur.551...67P} {551, 67}

\bibitem[\protect\citeauthoryear{{Planck Collaboration} et~al.,}{{Planck
  Collaboration} et~al.}{2016}]{2016A&A...594A..13P}
{Planck Collaboration} et~al., 2016, \mn@doi [\aap]
  {10.1051/0004-6361/201525830}, \href
  {http://cdsads.u-strasbg.fr/abs/2016A%26A...594A..13P} {594, A13}

\bibitem[\protect\citeauthoryear{{Radovich} et~al.,}{{Radovich}
  et~al.}{2004}]{Radovichetal2004}
{Radovich} M.,  et~al., 2004, \mn@doi [\aap] {10.1051/0004-6361:20034458},
  \href {http://cdsads.u-strasbg.fr/abs/2004A%26A...417...51R} {417, 51}

\bibitem[\protect\citeauthoryear{{Rodriguez}, {Chatterjee}  \&
  {Rasio}}{{Rodriguez} et~al.}{2016}]{2016PhRvD..93h4029R}
{Rodriguez} C.~L.,  {Chatterjee} S.,   {Rasio} F.~A.,  2016, \mn@doi [\prd]
  {10.1103/PhysRevD.93.084029}, \href
  {https://ui.adsabs.harvard.edu/abs/2016PhRvD..93h4029R} {93, 084029}

\bibitem[\protect\citeauthoryear{{Schutz}}{{Schutz}}{1986}]{1986Natur.323..310S}
{Schutz} B.~F.,  1986, \mn@doi [\nat] {10.1038/323310a0}, \href
  {https://ui.adsabs.harvard.edu/abs/1986Natur.323..310S} {323, 310}

\bibitem[\protect\citeauthoryear{{Singer} \& {Price}}{{Singer} \&
  {Price}}{2016}]{2016PhRvD..93b4013S}
{Singer} L.~P.,  {Price} L.~R.,  2016, \mn@doi [\prd]
  {10.1103/PhysRevD.93.024013}, \href
  {http://cdsads.u-strasbg.fr/abs/2016PhRvD..93b4013S} {93, 024013}

\bibitem[\protect\citeauthoryear{{Skrutskie} et~al.,}{{Skrutskie}
  et~al.}{2006}]{Skrutskieetal2006}
{Skrutskie} M.~F.,  et~al., 2006, \mn@doi [\aj] {10.1086/498708}, \href
  {http://cdsads.u-strasbg.fr/abs/2006AJ....131.1163S} {131, 1163}

\bibitem[\protect\citeauthoryear{{Stone}, {Metzger}  \& {Haiman}}{{Stone}
  et~al.}{2017}]{2017MNRAS.464..946S}
{Stone} N.~C.,  {Metzger} B.~D.,   {Haiman} Z.,  2017, \mn@doi [\mnras]
  {10.1093/mnras/stw2260}, \href
  {http://cdsads.u-strasbg.fr/abs/2017MNRAS.464..946S} {464, 946}

\bibitem[\protect\citeauthoryear{{Strader}, {Chomiuk}  \& {Tremou}}{{Strader}
  et~al.}{2017}]{2017TNSCR.833....1S}
{Strader} J.,  {Chomiuk} L.,   {Tremou} E. e.~a.,  2017, Transient Name Server
  Classification Report, \href
  {http://cdsads.u-strasbg.fr/abs/2017TNSCR.833....1S} {833}

\bibitem[\protect\citeauthoryear{{The LIGO Scientific Collaboration} \& {the
  Virgo Collaboration}}{{The LIGO Scientific Collaboration} \& {the Virgo
  Collaboration}}{2017a}]{GCN21474}
{The LIGO Scientific Collaboration} {the Virgo Collaboration} 2017a, GRB
  Coordinates Network, Circular Service, No.~21474, \#1 (2017a), 21474

\bibitem[\protect\citeauthoryear{{The LIGO Scientific Collaboration} \& {the
  Virgo Collaboration}}{{The LIGO Scientific Collaboration} \& {the Virgo
  Collaboration}}{2017b}]{GCN21493}
{The LIGO Scientific Collaboration} {the Virgo Collaboration} 2017b, GRB
  Coordinates Network, Circular Service, No.~21493, \#1 (2017b), 21493

\bibitem[\protect\citeauthoryear{{The LIGO Scientific Collaboration} \& {the
  Virgo Collaboration}}{{The LIGO Scientific Collaboration} \& {the Virgo
  Collaboration}}{2017c}]{GCN21934}
{The LIGO Scientific Collaboration} {the Virgo Collaboration} 2017c, GRB
  Coordinates Network, Circular Service, No.~21934, \#1 (2017c), 21934

\bibitem[\protect\citeauthoryear{{The LIGO Scientific Collaboration} \& {the
  Virgo Collaboration}}{{The LIGO Scientific Collaboration} \& {the Virgo
  Collaboration}}{2019}]{2019arXiv190103310T}
{The LIGO Scientific Collaboration} {the Virgo Collaboration} 2019, arXiv
  1901.03310, \href {http://cdsads.u-strasbg.fr/abs/2019arXiv190103310T} {}

\bibitem[\protect\citeauthoryear{{The LIGO Scientific Collaboration}
  et~al.,}{{The LIGO Scientific Collaboration}
  et~al.}{2018b}]{2018arXiv181112940T}
{The LIGO Scientific Collaboration} et~al., 2018b, arXiv 181112940T, \href
  {http://cdsads.u-strasbg.fr/abs/2018arXiv181112940T} {}

\bibitem[\protect\citeauthoryear{{The LIGO Scientific Collaboration}
  et~al.,}{{The LIGO Scientific Collaboration}
  et~al.}{2018a}]{2018arXiv181112907T}
{The LIGO Scientific Collaboration} et~al., 2018a, arXiv 181112907T, \href
  {http://adsabs.harvard.edu/abs/2018arXiv181112907T} {}

\bibitem[\protect\citeauthoryear{{Veitch} et~al.,}{{Veitch}
  et~al.}{2015}]{2015PhRvD..91d2003V}
{Veitch} J.,  et~al., 2015, \mn@doi [\prd] {10.1103/PhysRevD.91.042003}, \href
  {http://cdsads.u-strasbg.fr/abs/2015PhRvD..91d2003V} {91, 042003}

\bibitem[\protect\citeauthoryear{{Wenger} et~al.,}{{Wenger}
  et~al.}{2000}]{2000A&AS..143....9W}
{Wenger} M.,  et~al., 2000, \mn@doi [\aaps] {10.1051/aas:2000332}, \href
  {http://cdsads.u-strasbg.fr/abs/2000A%26AS..143....9W} {143, 9}

\bibitem[\protect\citeauthoryear{{Yamazaki}, {Asano}  \& {Ohira}}{{Yamazaki}
  et~al.}{2016}]{2016PTEP.2016e1E01Y}
{Yamazaki} R.,  {Asano} K.,   {Ohira} Y.,  2016, \mn@doi [Progress of
  Theoretical and Experimental Physics] {10.1093/ptep/ptw042}, \href
  {http://cdsads.u-strasbg.fr/abs/2016PTEP.2016e1E01Y} {2016, 051E01}

\bibitem[\protect\citeauthoryear{{Zhang}}{{Zhang}}{2016}]{2016ApJ...827L..31Z}
{Zhang} B.,  2016, \mn@doi [\apjl] {10.3847/2041-8205/827/2/L31}, \href
  {http://adsabs.harvard.edu/abs/2016ApJ...827L..31Z} {827, L31}

\bibitem[\protect\citeauthoryear{{de Mink} \& {King}}{{de Mink} \&
  {King}}{2017}]{deMink&King2017}
{de Mink} S.~E.,  {King} A.,  2017, \mn@doi [\apjl] {10.3847/2041-8213/aa67f3},
  \href {http://cdsads.u-strasbg.fr/abs/2017ApJ...839L...7D} {839, L7}

\makeatother
\end{thebibliography}

\end{document}